\documentclass[12pt]{article}
\usepackage{psfig,epsf,amssymb,amsmath,latexsym}
\title{Perturbation theory for Lyapunov exponents \\
of an Anderson model on a strip}

\author{Hermann Schulz-Baldes
\\
\\
{\small Institut f\"ur Mathematik, Technische Universit{\"a}t Berlin, 
10623 Berlin, Germany}
\\
\\
}

\date{ }

\newtheorem{theo}{Theorem}

\newtheorem{proposi}{Proposition}
\newtheorem{lemma}{Lemma}
\newtheorem{coro}{Corollary}

\newcommand{\CC}{{\mathbb C}}

\newcommand{\RR}{{\mathbb R}}

\newcommand{\ZZ}{{\mathbb Z}}

\newcommand{\EE}{{\bf E}}

\newcommand{\Ee}{{\cal E}}
\newcommand{\Ff}{{\cal F}}

\newcommand{\Oo}{{\cal O}}

\newcommand{\Uu}{{\cal U}}
\newcommand{\lap}{{\Delta_{L}}}
\newcommand{\rw}{{{\mbox{\rm w}}}}
\newcommand{\av}{{\mbox{\rm\tiny av}}}

\begin{document}

\maketitle

\begin{abstract}
It is proven that the inverse 
localization length of an Anderson model on a
strip of width $L$ is bounded above by $L/\lambda^2$ for small values
of the coupling constant  $\lambda$ of the disordered potential. For
this purpose, a formalism is developed in order to calculate the
bottom Lyapunov exponent  associated with random products of large
symplectic matrices perturbatively in the coupling constant of the
randomness.
\end{abstract}

\vspace{.5cm}

\section{Main result and discussion}
\label{sec-results}

The Anderson model describes electronic waves scattered at random
obstacles. Here the physical space is supposed to be
quasi one dimensional and given by
an infinite strip of finite but large width $L$.
Hence the Hilbert space is
$\ell^2(\ZZ,\CC^L)$ and states therein will be decomposed as  
$\psi=(\psi(n))_{n\in\ZZ}$ with $\psi(n)\in\CC^L$.
The Anderson Hamiltonian on a strip is then defined by

$$
(H_L(\lambda)\psi)(n)
\;=\;
-\psi(n+1)-\psi(n-1)+\lap\psi(n)+\lambda V(n)\psi(n)
\mbox{ . }
$$

\noindent Here $\lap:\CC^L\to\CC^L$ is the transverse (one
dimensional) discrete laplacian with
periodic boundary conditions; denoting the cyclic shift on $\CC^L$ by $S$,
it is given by $\lap=-S-S^*$. For $L=1,2$, one may rather set
$\Delta_1=0$ and $\Delta_2=-S=-S^*$, but our main interest will be in
the case $L\geq 3$ anyway. Furthermore,
$\lambda\in\RR$ is the coupling constant
of the random potential $V(n):\CC^L\to\CC^L$ which is a diagonal matrix
$V(n)=\mbox{diag}(v(n,1),\ldots ,v(n,L))$. All real numbers
$(v(n,l))_{n\in\ZZ,\,l=1,\ldots,L}$ are independent and identically
distributed centered real variables with unit variance.
Given a fixed energy $E\in\RR$, it is convenient to rewrite the eigenvalue
equation $H_L(\lambda)\psi=E\psi$ using the transfer matrices

\begin{equation}
\label{eq-transferdef}
\left(\begin{array}{c}
\psi(n+1) \\ \psi(n) 
\end{array}
\right)
\;=\;
T(n)
\left(\begin{array}{c}
\psi(n) \\ \psi(n-1) 
\end{array}
\right)
\mbox{ , }
\qquad
T(n)
\;=\;
\left(\begin{array}{cc} 
\lap + \lambda V(n) - E\,{\bf 1}  & - {\bf 1}
\\
{\bf 1} & 0 
\end{array}
\right)
\mbox{ . } 
\end{equation}

This work concerns the study of
the asymptotics of their random products which is characterized by the
Lyapunov exponents. One way \cite{BL,CL}
to define these exponents is to use a
formalism of second quantization. For $p=1,\ldots,L$, let
$\Lambda^p\CC^{2L}$ denote the Hilbert space of the anti-symmetrized
$p$-fold tensor products of $\CC^{2L}$, the scalar product being given
via the determinant. Given a linear map $T$ on 
$\CC^{2L}$, its second quantized
$\Lambda^p T$ on $\Lambda^p\CC^{2L}$ is defined as
usual. Now the whole family of non-negative 
Lyapunov exponents $\gamma_1\geq \gamma_2\geq\ldots\geq \gamma_L\geq
0$ are defined by:

\begin{equation}
\label{eq-lyapform0}
\sum_{l=1}^p\,\gamma_l
\;=\;
\lim_{N\to\infty}\,\frac{1}{N}\;\EE\,\log
\left(\,\left\|\;\prod_{n=1}^N\Lambda^pT(n)\;
\right\|\right)
\mbox{ , }
\qquad
p\;=\;1,\ldots,L
\mbox{ , }
\end{equation}

\noindent where the expectation is taken over all random variables. 
Positivity of the bottom Lyapunov exponent $\gamma_L$ was already
known \cite{BL,GM,GR} and is sufficient to assure Anderson 
localization \cite{KLS,BL}. The object of this work is to provide a
quantitative lower bound for small coupling of the random potential. 

\vspace{.2cm}

We will need a hypothesis excluding exceptional energies with 
Kappus-Wegner breakdown of the
perturbation theory to leading order \cite{KW}. This can
in principle be overcome \cite{KW,BK,CK}. Let $[c]$ denote the integer part
of $c\in\RR$.
 
\vspace{.2cm}

\noindent {\bf Main hypothesis:} 
Let $\mu_l=-2\,\cos(\frac{2\pi l}{L})-E$ and, if 
$|\mu_l|\leq 2$,  set
$e^{\imath\eta_l}=\frac{1}{2}(\mu_l+\imath\sqrt{4-\mu_l^2})$.
Except if $\sigma=-1$ and $j=k$ or $\{k,l\}=\{m,j\}$, we suppose

\begin{equation}
\label{eq-hypothesis}
e^{\imath(\eta_k+\sigma\eta_j)}
\;\neq\;
1
\mbox{ , }
\qquad
e^{\imath(\eta_k+\eta_l+\sigma\eta_m+\sigma\eta_j)}
\;\neq\;
1
\mbox{ , }
\qquad
k,l,m,j=1,\ldots,\left[\frac{L+1}{2}\right]
\mbox{ , }
\;\;\sigma=\pm 1
\mbox{ . }
\end{equation}

\begin{theo} 
\label{theo-main} Suppose $E\in\RR$ is in the spectrum of $H_L(0)$ and
satisfies the Main hypothesis.

\vspace{.1cm}

\noindent
{\rm (i)} Then

$$
\gamma_L
\;\geq\;
\frac{\lambda^2}{8L}\;+\;\Oo(\lambda^3)
\mbox{ . }
$$

\noindent {\rm (ii)} Let $E_b$ be a band edge of $H_L(0)$ and
$E=E_b+\epsilon$ be in the spectrum of $H_L(0)$. Then 

$$
\gamma_L
\;\geq\;
\frac{\lambda^2}{8L}\;
\frac{1}{|\epsilon|}\;+\;\Oo(\lambda^3)
\mbox{ . }
$$

\end{theo}

\vspace{.2cm}

For the case $L=1$, this was proven by Pastur and Figotin 
\cite{PF} and in a related situation of hamiltonian 
stochastic differential equations by
Arnold, Papanicolaou and Wihstutz \cite{APW}. 
Actually, we go beyond the above theorem and prove an asymptotic formula for
$\gamma_L$ in Theorem \ref{theo-Lyapasymp} in Section 
\ref{sec-perturbative} below. 
We then argue (non-rigorously) in Section \ref{sec-speculation}
that the bound (i) gives the right
order of magnitude for all energies away from the band edges and the
so-called internal band edges which are
defined by the property that $\eta_l=0$ for
some $l$. Near the band edges, 
the bottom Lyapunov exponent is much larger according to
item (ii). Indications for such stronger localization properties
in this regime appeared also in \cite{Klo}.
It is straightforward to analyse the
large deviations of the growth behavior of the transfer matrices 
around the typical behavior given by the Lyapunov exponent
with the techniques of \cite[Section 5]{JSS}.
If one adds the supplementary hypothesis that
$\EE(v(n,l)^3)=0$, then the corrections are 
actually of the order $\Oo(\lambda^4)$.
The main deficiency of the present work is the lack of control of the
error term on the strip with $L$ and the energy $E$.

\vspace{.2cm}

The method of proof transposes directly 
to a more abstract setting of random products of symplectic matrices 
with small coupling, if only one supposes that the modulus one 
eigenvalues of the
unperturbed part are non-degenerate (in the language of Section
\ref{sec-channels}, elliptic
channels are then non-degenerate).
In order to deal with the degeneracies appearing in the Anderson model,
the concrete form of the random perturbation in (\ref{eq-transferdef})
is however heavily used.
We believe that hamiltonian stochastic differential equations could also be
treated. Preliminary results in this framework
were obtained by Teichert \cite{Tei}.

\vspace{.2cm}

Let us briefly describe the key steps of the proof. First the transfer
matrix at $\lambda=0$ is diagonalized into symplectic blocks given by
rotations (Sections \ref{sec-normal} and \ref{sec-coeff}). Then the matrix
elements of the random perturbation are calculated
in that representation (Section
\ref{sec-perturb}). This normal form allows to derive
a basic perturbative formula for the Lyapunov exponents 
(Sections \ref{sec-lyapsymp} and \ref{sec-perturbative}).
A new ingredient herein is the consistent use of
symplectic frames. It is then possible to apply
a crucial identity related to the geometry of Lagrangian manifolds (Lemma
\ref{lem-lagrangian} in Section \ref{sec-unitaries}).
The normal form of the transfer matrix now
allows to efficiently control the
oscillatory sums appearing in the perturbative formula 
(Section \ref{sec-osci}). There
is an inessential technical difficulty due to the presence of
so-called hyperbolic channels. They do not appear if $L$ is odd and
$E$ is near the band center. The text is written such that the reader
can understand this case and
hence the main point of the argument by skipping Section
\ref{sec-separ} and then omitting Sections \ref{sec-ellweights} and 
\ref{sec-perturbative}.

\vspace{.2cm}

In order to compare Theorem \ref{theo-main}
with results in the physics literature, 
let us interprete $L$ as the number of
channels of a disordered wire and $\gamma_L$ as the associated 
inverse localization length. Then the behaviour

\begin{equation}
\gamma_L\;\sim\;
\frac{\lambda^2}{L}\;+\;\Oo(\lambda^3)
\mbox{  }
\label{eq-behave}
\end{equation}

\noindent confirms the predictions of Thouless
\cite{Tho} as well as the Dorokhov-Mello-Pereyra-Kumar theory (see
\cite{Ben} for a review on the latter). 

\vspace{.2cm}

For the Anderson model in two dimensions, all waves are expected to 
localize even at small disorder 
\cite{AALR}. Few rigorous results indicating such a phenomenon are known. 
Even though one may think of the strip as an approximation to the
two dimensional situation, it is unlikely that (\ref{eq-behave})
gives much insight. 
For a proof of localization, one would need to
prove the so-called initial length scale estimate in order to apply the
multiscale analysis \cite{FS,DK,GK}. It states that the wave functions on
a square of appropriate diameter decrease from center to boundary (or
inversely) with a high probability.
But even if the error term in (\ref{eq-behave}) could be
neglected at small $\lambda$, exponential decay of typical
eigenfunctions is noticeable only on a length scale $N$ given by
$N\gamma_L=\Oo(1)$, that is $N=L/\lambda^2$, which is much
larger than the strip width $L$. Therefore
(\ref{eq-behave}) is of interest only in the quasi one dimensional
situation. 
Indeed, Anderson localization in two
dimensions is expected to be a non-perturbative phenomenon
(like BCS
theory) and thus not tractable by a ``naive perturbation theory'' as
developed here. This is reflected by the prediction that the 2D localization
length 
behaves non-analytically like $e^{1/\lambda^2}$ for small
$\lambda$ \cite{AALR}. Rigorously known is only a lower bound on
the phase-space localization of the eigenfunctions \cite{SSW}.

\vspace{.2cm}

\section{Analysis of the transfer matrix}
\label{sec-transfer}

Each transfer matrix $T(n)$ is a random element
of the symplectic group 

$$
\mbox{SP}(2L,\RR)
\;=\;
\left\{ \,T\in \mbox{M}_{2L\times 2L}(\RR)\;\left|\;
T^tJT=J\;\right.\right\}
\mbox{ , }
\qquad
J
\;=\;
\left(\begin{array}{cc} 
0 & - {\bf 1}
\\
{\bf 1} & 0 
\end{array}
\right)
\mbox{ . } 
$$

\noindent The aim of this section is to construct a symplectic 
basis transformation $M\in\,$SP$(2L,\RR)$ such that

\begin{equation}
\label{eq-normal}
M^{-1}\, T(n)\,M
\;=\;
R\,({\bf 1}-\lambda\, P(n))
\mbox{ , }
\end{equation}

\noindent where the free transfer matrix $R$ 
({\sl i.e.} $\lambda=0$) takes a particularly
simple form given by a direct sum of rotations. 
The random perturbation $P(n)$ lies in the Lie algebra sp$(2L,\RR)$.
Its matrix elements and some of their expectation values will be
calculated below. Throughout this section the index $n$ is
kept fixed and will thus be suppressed.

\subsection{Normal form without disorder}
\label{sec-normal}

Let us introduce, for $l=1,\ldots,L$,

$$
\phi_l=\left(\begin{array}{c}\phi_l(1) \\ \vdots \\ \phi_l(L)
\end{array}\right)
\in\CC^L
\mbox{ , }
\qquad
\phi_l(k)\;=\;\frac{1}{\sqrt{L}}\;
\exp\left(\frac{2\pi\imath\,lk}{L}\right) 
\mbox{ . }
$$

\noindent Then $\lap\phi_l=-2\cos(2\pi\,l/L)\,\phi_l$.
Note that the fundamental
$\Phi_L=\phi_L$ is real; moreover, for even $L$, the vector 
$\Phi_{L/2}=\phi_{L/2}$ is real as well. For other $l$,
real normalized eigenvectors are obtained by

$$
(\Phi_l,\Phi_{L-l})
\;=\;
(\phi_l,\phi_{L-l})\;\frac{1}{\sqrt{2}}\,
\left(
\begin{array}{cc}
-\imath & 1 \\
\imath & 1
\end{array}
\right)
\mbox{ . }
$$

\noindent Next define an orthogonal matrix $m\in\,$O$(L,\RR)$ and
unitaries $d,f\in\, $U$(L,\CC)$ by

$$
m\;=\;(\Phi_1,\ldots,\Phi_L)
\mbox{ , }
\qquad
f\;=\;(\phi_1,\ldots,\phi_L)
\mbox{ , }
\qquad
m\;=\;f\,d
\mbox{ . }
$$

\noindent Finally introduce the diagonal matrix
$\mu=\mbox{diag}(\mu_1,\ldots,\mu_L)$ where $\mu_l=-2\cos(2\pi\,l/L)-E$.
With these notations,

$$
m^*(\lap-E)m
\;=\;
f^*(\lap-E)f
\;=\;
\mu
\mbox{ . }
$$

\vspace{.2cm}

An eigenvalue $\mu_l$ will be
called elliptic if $|\mu_l|<2$, hyperbolic if $|\mu_l|>2$ and
parabolic if $|\mu_l|=2$. Here energies $E$ for which
there are parabolic eigenvalues are excluded by the hypothesis
(\ref{eq-hypothesis}).  An energy $E$ is in the spectrum of $H_L(0)$,
the Laplacian on $\ell^2(\ZZ,\CC^L)$ if there exists an elliptic or a
parabolic eigenvalue. The spectrum of $H_L(0)$ is hence $[-4,4]$ if
$L>2$ is even and $[-4,2+2\,\cos(\pi/L)]$ if $L>1$ is odd.
If $L$ is odd and $E$ slightly above the band center (Fig.~1(ii)), 
all eigenvalues are elliptic while for
energies $E$ outside of
the spectrum, all eigenvalues
are hyperbolic. In between one has both hyperbolic and elliptic
eigenvalues. 
For notational convenience, we will suppose that $\mu_l<2$ for all
$l$. Hence there exists $L_h\leq
\frac{L}{2}$ such that
that $\mu_l$ is hyperbolic for $l=0,\ldots,L_h$, and elliptic for
$l=L_h+1,\ldots, [\frac{L}{2}]$.
Moreover, there is a degeneracy $\mu_{L-l}=\mu_l$ due to reflection
symmetry  which will be
further analyzed in Section \ref{sec-channels}.
In case there are no hyperbolic eigenvalues, 
let us set $L_h=-1$. In Fig.~1 are shown
examples of (i) a situation with mixed elliptic and hyperbolic
eigenvalues and (ii) a situation with only elliptic ones.

\begin{figure}
\centerline{\psfig{figure=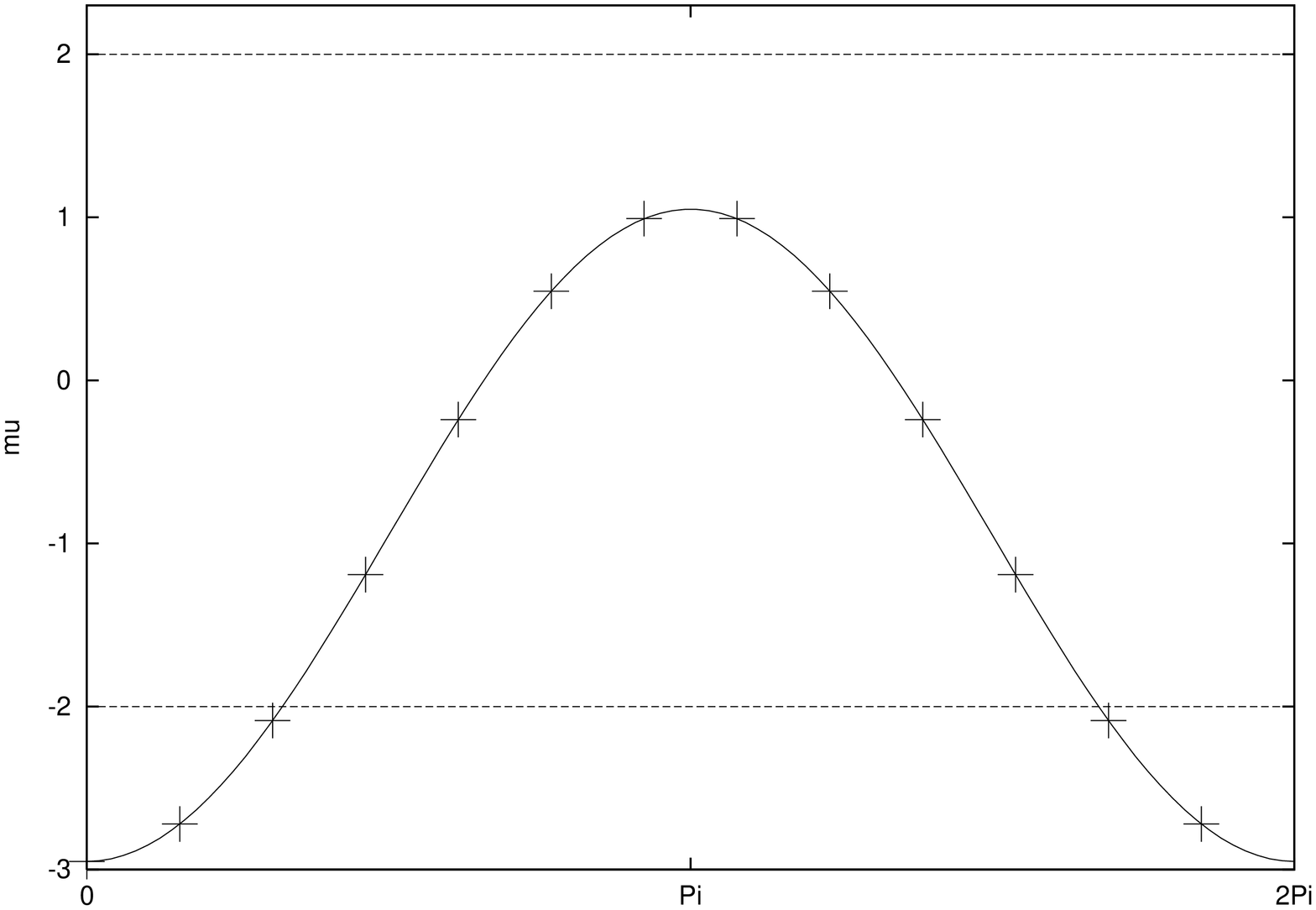,height=6.6cm,width=6.6cm,angle=270}
$\;\;\;\;\;\;\;\;\;\;\;$
\psfig{figure=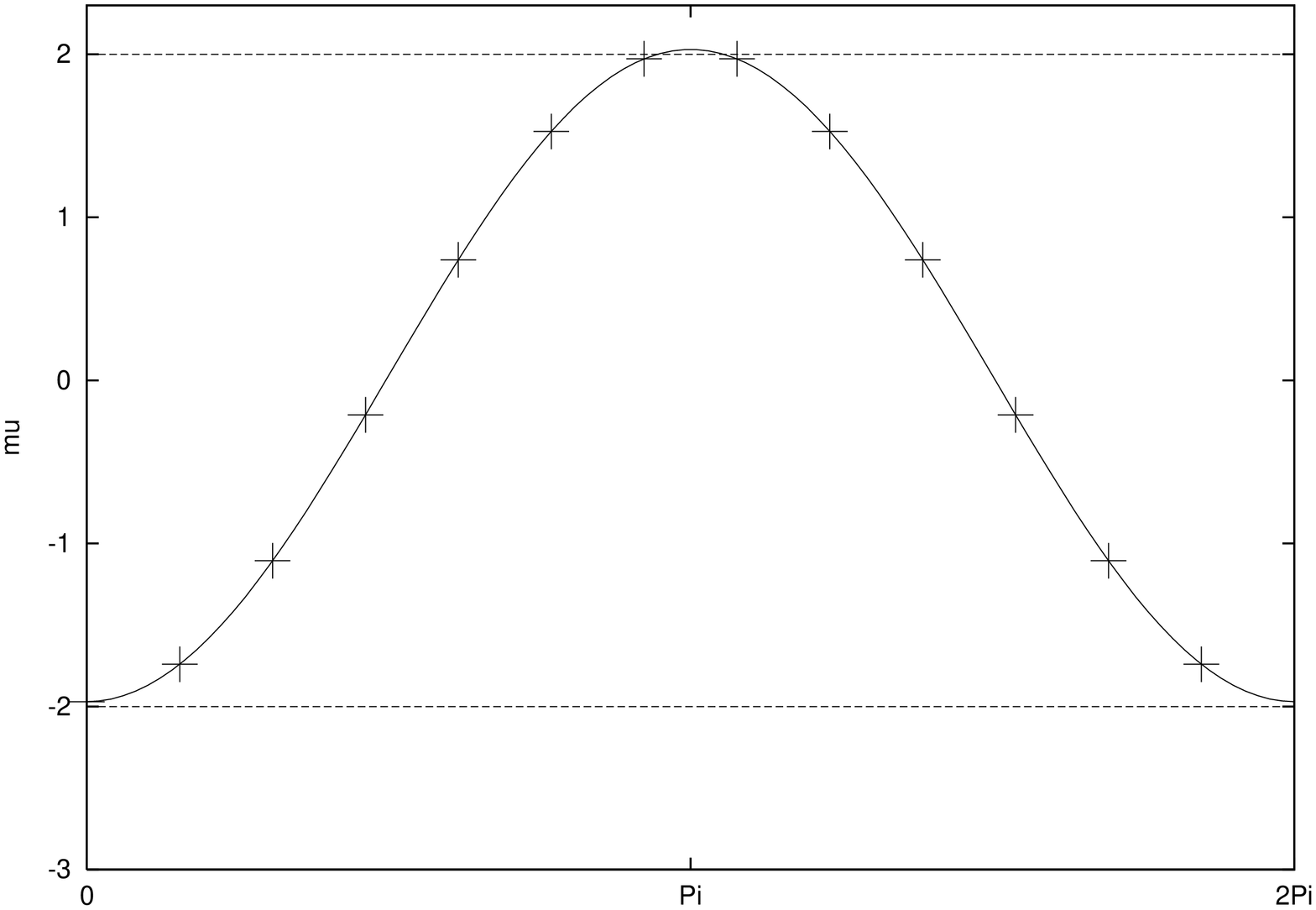,height=6.6cm,width=6.6cm,angle=270}}
\caption{{\sl Plot of the the energy levels for $L=13$. {\rm (i)}
Here $E=0.95$ and $L_h=2$. 
The eigenvalues $\mu_0$, $\mu_1$, $\mu_2$, $\mu_{11}$ and
$\mu_{12}$ are outside of the window $[-2,2]$ and hence hyperbolic. 
{\rm (ii)} Here $E=-0.03$. All eigenvalues are elliptic.
}} 
\end{figure}

\vspace{.2cm}

For later use, let us
set $g=$ diag$(g_1,\ldots,g_L)$ where $g_l=1$ if
$\mu_l$ is elliptic and $g_l=\imath$ if $\mu_l$ is hyperbolic.
Note that $[d,\mu]=[d,g]=[\mu,g]=0$. 

\vspace{.2cm}

In order to diagonalize the transfer matrix, let us introduce the
diagonal complex $L\times L$ matrix 
$\kappa=\frac{\mu}{2}+\sqrt{\frac{\mu^2}{4}-1}$ as
well as the (possibly complex-valued) matrices

$$
N
\;=\;
\left(
\begin{array}{cc}
m\,\sqrt{\frac{1}{{\kappa-\frac{1}{\kappa}}}} &
m\,\sqrt{\frac{1}{{\kappa-\frac{1}{\kappa}}}} \\
& \\
m\,\sqrt{\frac{1}{{\kappa-\frac{1}{\kappa}}}}\,\frac{1}{\kappa} &
m\,\sqrt{\frac{1}{{\kappa-\frac{1}{\kappa}}}}\,\kappa 
\end{array}
\right)
\mbox{ , }
\qquad
N^{-1}
\;=\;
\left(
\begin{array}{cc}
\sqrt{\frac{1}{{\kappa-\frac{1}{\kappa}}}}\,\kappa\,m^* &
-\sqrt{\frac{1}{{\kappa-\frac{1}{\kappa}}}}\,m^* \\
& \\
-\sqrt{\frac{1}{{\kappa-\frac{1}{\kappa}}}}\,\frac{1}{\kappa} \,m^*&
\sqrt{\frac{1}{{\kappa-\frac{1}{\kappa}}}}\,m^*
\end{array}
\right)
\mbox{ , }
$$

\noindent where here and below all roots are taken on the first branch.
Then one immediately verifies that

$$
N^{-1}\,\left(
\begin{array}{cc}
\lap-E & -{\bf 1} \\
{\bf 1} & 0
\end{array}
\right)\,N
\;=\;
\left(
\begin{array}{cc}
\kappa & 0 \\
0 & \frac{1}{\kappa}
\end{array}
\right)
\mbox{ . }
$$

\vspace{.2cm}

Finally the r.h.s. will be transformed into a normal form which is
real-valued and symplectic. For elliptic $\mu_l$ define $0<\eta_l<\pi$ by
$\kappa_l=e^{\imath \eta_l}$, for a hyperbolic one $\eta_l>0$ by
$\kappa_l=e^{\eta_l}$. Then
introduce the rotation and hyperbolic rotation matrices by
($\eta\in\RR$): 

$$
R_e(\eta)
\;=\;
\left(
\begin{array}{cc}
\cos(\eta) & -\sin(\eta) \\
\sin(\eta) & \cos(\eta)
\end{array}
\right)
\mbox{ , }
\qquad
R_h(\eta)
\;=\;
\left(
\begin{array}{cc}
\cosh(\eta) & \sinh(\eta) \\
\sinh(\eta) & \cosh(\eta)
\end{array}
\right)
\mbox{ . }
$$

\noindent Setting

$$
C
\;=\;
\sqrt{\frac{\imath}{2}}\;
\left(
\begin{array}{cc}
{\bf 1} & \imath\,{\bf 1} \\
{\bf 1} & -\imath\,{\bf 1}
\end{array}
\right)
\mbox{ , }
\qquad
G
\;=\;
\left(
\begin{array}{cc}
\overline{g^{\frac{1}{2}}} & 0 \\
0 & g^2\,g^{\frac{1}{2}}
\end{array}
\right)
\mbox{ , }
$$

\noindent one verifies

$$
G^{-1} \,C^{-1}\,\left(
\begin{array}{cc}
\kappa & 0 \\
0 & \frac{1}{\kappa}
\end{array}
\right)
\,C\, G
\;=\;
R_1(\eta_1)
\oplus\ldots\oplus R_L(\eta_L)
\mbox{ , }
$$

\noindent where $R_l(\eta_l)$ is either $R_e(\eta_l)$ or
$R_h(\eta_l)$ depending on whether $\mu_l$ is elliptic or hyperbolic,
and where the direct sum is understood such that $R_l(\eta_l)$ acts on
the $l$th and $(l+L)$th component of $\CC^{2L}$, namely the normal form
$R=R_1(\eta_1)
\oplus\ldots\oplus R_L(\eta_L)$ is a real-valued symplectic matrix.
In case that there are parabolic eigenvalues, the normal form contains
Jordan blocs just as in \cite{SB}.

\vspace{.2cm}

Let us resume the above. Set $M=NCG$, then

$$
M^{-1}\,\left(
\begin{array}{cc}
\lap-E & -{\bf 1} \\
{\bf 1} & 0
\end{array}
\right)\,M
\;=\;
R
\mbox{ . }
$$

\noindent Furthermore the symplectic basis change $M\in\,$SP$(2L,\RR)$
is more explicitly given by

$$
M
\;=\;
\left(
\begin{array}{cc}
m\,h &
0 \\
& \\
m\,\frac{1}{2}\,(\kappa+\frac{1}{\kappa})\,h &
m\,h^{-1}\,g^2
\end{array}
\right)
\mbox{ , }
\qquad
M^{-1}
\;=\;
\left(
\begin{array}{cc}
h^{-1}\,m^* &
0 \\
& \\
-\frac{1}{2}\,g^2\,(\kappa+\frac{1}{\kappa})\,h\,m^* &
g^2\,h\,m^* 
\end{array}
\right)
\mbox{ , }
$$

\noindent where 

$$
h
\;=\;
\sqrt{\frac{2\imath\,g^3}{\kappa-\frac{1}{\kappa}}}
\;=\;
\mbox{diag}(h_1,\ldots,h_L)
\mbox{ , }
\qquad
h_l
\;=\;
\left\{
\begin{array}{cc}
\sin(\eta_l)^{-\frac{1}{2}}
&
\mu_l \;\mbox{ elliptic,}
\\
& \\
\sinh(\eta_l)^{-\frac{1}{2}}
&
\;\;\;\;\mu_l \;\mbox{ hyperbolic.}
\end{array}
\right.
$$

\subsection{Calculation of the perturbation}
\label{sec-perturb}

First note that

$$
T
\;=\;
\left(
\begin{array}{cc}
\lap-E & -{\bf 1} \\
{\bf 1} & 0
\end{array}
\right)\;
\left(\,
\left(
\begin{array}{cc}
{\bf 1} & 0 \\
0 & {\bf 1}
\end{array}
\right)\;-\,\lambda\,
\left(
\begin{array}{cc}
0 & 0 \\
V & 0
\end{array}
\right)
\,\right)
\mbox{ . }
$$

\noindent Hence $P\in\,$sp$(2L,\RR)$ 
in equation (\ref{eq-normal}) is given by

\begin{equation}
P
\;=\;
M^{-1}\;
\left(
\begin{array}{cc}
0 & 0 \\
V & 0
\end{array}
\right)
M
\;=\;
\left(
\begin{array}{cc}
0 &  0 \\
g^2\,h\,d^*\,\hat{V}
\,d\,h & 0
\end{array}
\right)
\mbox{ . }
\label{eq-Pnormal}
\end{equation}

\noindent Here the identity $m=fd$ was used as well as the definition
$\hat{V}= f^*Vf$. Actually $f$ is the matrix of the discrete
Fourier transform so that $\hat{V}=\hat{V}^*$ is the Toeplitz matrix
associated with the Fourier transform of the random potential (at fixed 
height $n$). More precisely, define

$$
\hat{v}(k)
\;=\;
\frac{1}{L}\;\sum_{l=1}^L\,v(l)\,
\exp\left(\frac{2\pi\imath\,lk}{L}\right)
\mbox{ . }
$$

\noindent One has $\overline{\hat{v}(k)}=\hat{v}(-k)$ and 
$\hat{v}(k+L)=\hat{v}(k)$. Then 

$$
\hat{V}
\;=\;
\left(
\begin{array}{cccc}
\hat{v}(0) & \hat{v}(1) & \ldots & \hat{v}(L-1) \\
\hat{v}(-1) &  \ddots & \ddots & \vdots \\
\vdots & \ddots & \ddots & \hat{v}(1) \\
\hat{v}(-L+1) &  \ldots & \hat{v}(-1) & \hat{v}(0) \\
\end{array}
\right)
\mbox{ . }
$$

\subsection{Symplectic channels}
\label{sec-channels}

The symplectic channels of a symplectic matrix
are by definition the eigenspaces of its (possibly degenerate)
eigenvalue pairs $(\kappa,\frac{1}{\kappa})$. For the
elliptic-hyperbolic rotation $R=R_1(\eta_1)
\oplus\ldots\oplus R_L(\eta_L)$, they are the
maximal subspaces of $\CC^{2L}$
characterized by the property that $R$
rotates by the same angle. For the Anderson model on a strip studied
here, there are $L_c+1=[\frac{L}{2}]+1$ channels.
The $0$th channel is
associated with the fundamental $\mu_L$ and given
by the span of $\{e_L,e_{2L}\}$ where $e_l\in\CC^{2L}$ has
a non-vanshing entry equal to $1$ only in the $l$th component. 
If $L$ is even, also the
$\frac{L}{2}$th channel is simple and spanned by
$\{e_{L/2},e_{3L/2}\}$.
Due to the degeneracy 
$\mu_l=\mu_{L-l}$ of the spectrum of $\lap$, 
all other channels are doubly degenerate. 
For $l\neq 0,\frac{L}{2}$, the $l$th channel is spanned
by $\{e_l,e_{L-l},e_{l+L},e_{2L-l}\}$.
Let us denote the degeneracy of the $l$th channel by $\nu_l$.
In accordance with the above, the $l$th channel is called elliptic if
$|\mu_l|<2$ and hyperbolic if $|\mu_l|>2$.
With these notations, $\CC^{2L}$ is hence decomposed into a direct sum
of the $L_c+1$ channels.

\vspace{.2cm}

There are $L_h+1$
hyperbolic channels. As we supposed (for convenience) that $\mu_l<2$, the
hyperbolic channels are the first ones, namely channels
$0,\ldots,L_h$ (compare Fig.~1(i)). 

\vspace{.2cm}

The projection on the $l$th channel will be denoted by $\pi_l$. It
satisfies $[\pi_l,J]=0$.
The non-zero eigenvalues of $\pi_lR\pi_l$ are $e^{\pm\imath\eta_l}$ if
the $l$th channel is elliptic and $e^{\pm\eta_l}$ if it is
hyperbolic. Hence one can decompose $\pi_l$ into the corresponding
eigenspaces $\pi_l=\pi_l^++\pi_l^-$. Resuming all the above,

\begin{equation}
\label{eq-decomp}
{\bf 1}_{\CC^{2L}}
\;=\;
\sum_{l=0}^{L_c} \;\pi_l^++\pi_l^-
\mbox{ , }
\qquad
R\;\pi_l^\pm\;=\;e^{\pm\,\imath \,\overline{g_l}\,\eta_l}
\;\pi_l^\pm
\mbox{ . }
\end{equation}

\subsection{A privileged basis}
\label{sec-basis}

For explicit calculations in the next section, 
it will be convenient to dispose of 
a basis of $\CC^{2L}$ with the following properties: 

\begin{enumerate}

\item the basis vectors are orthonormal and compatible with the
symplectic structure;

\item the basis vectors are eigenvectors
of $R$;

\item the matrix elements of the perturbation $P$ w.r.t. the
basis are particularly simple, {\sl i.e.}, up to a constant, given by
the Fourier transform $\hat{v}$ of the potential. 

\end{enumerate}

For this purpose let us note that, for any $\eta\in\RR$, linearly independent
eigenvectors of $R_e(\eta)$
are $\left(\begin{array}{c} 1 \\ -\imath\end{array}\right)$ and 
$\left(\begin{array}{c} 1 \\ \imath\end{array}\right)$, 
while for $R_h(\eta)$ one can choose
$\left(\begin{array}{c} 1 \\ 1\end{array}\right)$ and 
$\left(\begin{array}{c} 1 \\ -1\end{array}\right)$. 
Comparing with equation (\ref{eq-Pnormal}), the third
property is verified if the top $L$ components of the basis vectors
commute with $h$ and are constructed with the inverse $d^*$ of $d$
or its quasi-inverse $d^t$. Indeed
it can be verified that $dd^t={\cal S}$ is the reflection in
$\CC^L$ sending component $l$ to component $(L-l)$. In
particular,
the components $\frac{L}{2}$ and $L$ are left invariant by
${\cal S}$. Moreover, ${\cal S}$ preserves the channels, namely
$[{\cal S},d]=[{\cal S},g]=[{\cal S},h]=0$ and
$\left(\begin{array}{cc} {\cal S}& 0 \\ 0 & {\cal S}
\end{array}\right)$ commutes with $\pi_l$.
Hence one is led to define the following:

\begin{equation}
\label{eq-basisdef}
W
\;=\;
\left(\rw^+_1,\ldots,\rw^+_L,\rw^-_1,\ldots,\rw^-_L\right)
\;=\;
\frac{1}{\sqrt{2}}\;
\left(
\begin{array}{cc}
d^* & d^t \\
-\imath g\,d^*  & \imath g\,d^t
\end{array}
\right)
\mbox{ . }
\end{equation}

\noindent It can readily be verified that 

\begin{equation}
\label{eq-basisid}
W^*W
\;=\;{\bf 1}
\mbox{ , }
\qquad
W^*\,J\,W
\;=\;
\frac{\imath}{2}
\left(
\begin{array}{cc}
g+g^* &  (g^*-g){\cal S} \\
(g-g^*){\cal S} & -g-g^* 
\end{array}
\right)
\mbox{ . }
\end{equation}

\noindent Hence the vectors
$|\rw^+_1\rangle,\ldots,|\rw^+_L\rangle,|\rw^-_1\rangle
,\ldots,|\rw^-_L\rangle$ defined by (\ref{eq-basisdef}) form
an orthonormal basis of $\CC^{2L}$ (we switch to Dirac notation for vectors
here).  
Let us write out examples for a simple and a double eigenvalue 
($\mu_L$ and $\mu_l$,  $1\leq l<\frac{L}{2}$, respectively)
more explicitly:

$$
|\rw^+_L\rangle
\;=\;
\frac{1}{\sqrt{2}}\,
\left(
|e_L\rangle-\imath g_L\,|e_{2L}\rangle
\right)
\mbox{ , }
\qquad
|\rw^+_l\rangle
\;=\;
\frac{1}{2}\,
(\imath\,|e_l\rangle+|e_{L-l}\rangle
+g_l\,|e_{L+l}\rangle-\imath g_l\,|e_{2L-l}\rangle)
\mbox{ . }
$$

\noindent Note that for an elliptic eigenvalue $\mu_l$,
$|{\rw}^-_l\rangle$ is the complex conjugate of 
$|\rw^+_l\rangle$, but if $\mu_l$ is hyperbolic, this only holds for the
first $L$ components.
As requested, one reads off from (\ref{eq-basisid}) that

\begin{equation}
\label{eq-basisJ}
J\,|\rw^\pm_l\rangle
\;=\;
\left\{
\begin{array}{cc}
\pm\,\imath |\rw^\pm_l\rangle & \mu_l \;\mbox{ elliptic , } \\
& \\
\mp\,|{\rw^\mp_{L-l}}\rangle & \;\;\;\;\;\;\mu_l \;\mbox{ hyperbolic . } 
\end{array}
\right.
\end{equation}

\noindent Also the second
required property follows from (\ref{eq-basisdef}):

$$
R\,|\rw^\pm_l\rangle
\;=\;
e^{\pm\imath\,\overline{g_l}\,\eta_l}\,|\rw^\pm_l\rangle
\mbox{ . }
$$

\noindent Finally, 

\begin{equation}
\label{eq-basisdecom}
\pi_0^\pm\;=\;|\rw^\pm_L\rangle\langle\rw^\pm_L|
\mbox{ , }
\qquad
\pi_l^\pm\;=\;|\rw^\pm_l\rangle\langle\rw^\pm_l|
+|\rw^\pm_{L-l}\rangle\langle\rw^\pm_{L-l}|
\mbox{ , }
\qquad
\pi_{\frac{L}{2}}^\pm\;=\;
|\rw^\pm_{\frac{L}{2}}\rangle\langle\rw^\pm_{\frac{L}{2}}|
\mbox{ . }
\end{equation}

\subsection{Matrix elements of the random perturbation}
\label{sec-coeff}

In this section, we show that the third desired property of the basis
stated in Section \ref{sec-basis}
is fulfilled and then exploit it in order to calculate the matrix
elements of the perturbation and some expectation values thereof.
Taking into account that $[d,h]=0$, it now follows from 
(\ref{eq-Pnormal}) and (\ref{eq-basisdef}) that

\begin{equation}
\label{eq-perturbcalc}
W^*PW
\;=\;
\frac{\imath}{2}
\,
\left(
\begin{array}{cc}
gh\hat{V}h & gh\hat{V}h{\cal S} \\
-{\cal S}gh\hat{V}h & -{\cal S}gh\hat{V}h {\cal S}\\
\end{array}
\right)
\mbox{ , }
\qquad
W^*P^*W
\;=\;
\frac{\imath}{2}
\,
\left(
\begin{array}{cc}
-h\hat{V}hg^* & h\hat{V}hg^*{\cal S} \\
-{\cal S}h\hat{V}hg^* & {\cal S}h\hat{V}hg^*{\cal S} \\
\end{array}
\right)
\mbox{ .}
\end{equation}

\noindent Hence one can read off, for two signs $\tau,\sigma$,

\begin{equation}
\label{eq-perturbform}
\langle{\rw}^\tau_l|P|{\rw}^\sigma_k\rangle
\;=\;
\tau\,\frac{\imath}{2}\,{g_l}\,h_lh_k\,\hat{v}(\sigma k-\tau l)
\mbox{ , }
\qquad
\langle{\rw}^\tau_l|P^*|{\rw}^\sigma_k\rangle
\;=\;
-\sigma\,\frac{\imath}{2}\,\overline{g_k}\,h_lh_k\,\hat{v}(\sigma k-\tau l)
\mbox{ . }
\end{equation}

\noindent Let us collect some useful identities. 

\begin{lemma}
\label{lem-matrixel}
Let $w_j^\pm\in\CC^{2L}$ be unit 
vectors satisfying $\pi_j^\pm
w_j^\pm=w_j^\pm$. Let $\tilde{P}=P+P^*$.

\vspace{.1cm}

\noindent {\rm (i)}  $\EE\,\langle w|P|w'\rangle=0$
for any $w,w'\in\CC^{2L}$.

\vspace{.1cm}

\noindent {\rm (ii)} If either $k\neq l$ or $\sigma\neq \tau$, then

$$
\EE\,
\langle w_l^\tau|\,P\,|w_k^\sigma\rangle
\;
\langle w_k^\sigma|\,P\,|w_l^\tau\rangle
\;=\;
-\tau\sigma\,\frac{1}{4L}\,
{g_l}\,{g_k}
\;h_l^2h_k^2
\mbox{ . }
$$

\vspace{.1cm}

\noindent {\rm (iii)} If either $k\neq l$ or $\sigma\neq \tau$, then

$$
\EE\,
\langle w_l^\tau|\,P^*\,|w_k^\sigma\rangle
\;
\langle w_k^\sigma|\,P\,|w_l^\tau\rangle
\;=\;
\frac{1}{4L}\;h_l^2h_k^2
\mbox{ . }
$$

\vspace{.1cm}

\noindent {\rm (iv)}
Let channels $l,k$ be elliptic. Then
$\pi_l^\sigma \tilde{P} \pi^\sigma_k=0$ and 
$\pi_l^\sigma \tilde{P} \pi^{-\sigma}_k=2\,\pi_l^\sigma {P} \pi^{-\sigma}_k$.
Moreover,

$$
\EE\,
\langle w_l^\sigma|\,\tilde{P}\,|w_k^{-\sigma}\rangle
\;
\langle w_k^{-\sigma}|\,\tilde{P}\,|w_l^\sigma\rangle
\;=\;
\frac{1}{L}
\;h_l^2h_k^2
\;=\;
\EE\,
\langle w_l^\sigma|\,\tilde{P}\,|w_k^{-\sigma}\rangle
\;
\langle \overline{w_l^\sigma}|\,\tilde{P}\,|\overline{w_k^{-\sigma}} \rangle
\mbox{ . }
$$

\vspace{.1cm}

\noindent {\rm (v)} 
$$
\EE\,\langle w_l^\sigma|\,|RP|^2\,|w_l^\sigma\rangle
\;=\;
\frac{1}{2}\;h_\av^2
\,h_l^2
\mbox{ , }
\qquad
\mbox{where}
\qquad
h_\av^2
\;=\;
\frac{1}{L}\,
\sum_{k=0}^{L_c}\nu_k\,h_k^2\,\cosh((1-g_k^2)\eta_k)
\mbox{ . }
$$

\end{lemma}

One might think of items (ii) through (v) as follows. Even though the
perturbation $P$ lifts the degeneracy  of the channels, taking
expectation values re-establishes it. 

\vspace{.2cm}

\noindent {\bf Proof.} {\rm (i)} This follows directly from
$\EE\,v(l)=0$ for all $l$.

\vspace{.1cm}

{\rm (ii)} Let 
$w^\tau_l=a_l\rw^\tau_l+b_l\rw^\tau_{L-l}$ with $|a_l|^2+|b_l|^2=1$ and
$b_l=0$ if the channel is simple, that is $l=0,\frac{L}{2}$. With these
notations, it follows from  (\ref{eq-perturbform}) 

\begin{eqnarray}
\langle w_l^\tau|\,P\,|w_k^\sigma\rangle
\;
\langle w_k^\sigma|\,P\,|w_l^\tau\rangle
& = &
-\sigma\tau\,\frac{1}{4}\,
{g_l}\,{g_k}
\,h_l^2h_k^2\,\cdot
\nonumber
\\
& &
\!\!\!\!\!\!\!\!\!\!\!\!\!\!\!
\cdot
\left(
\overline{a_l}{a_k}\hat{v}(\!\sigma k\!-\!\tau l)+
\overline{a_l}b_k\hat{v}(-\sigma k\!-\!\tau l)+
\overline{b_l}a_k\hat{v}(\!\sigma k\!+\!\tau l)+
\overline{b_l}{b_k}\hat{v}(-\sigma k\!+\!\tau l)
\right)\,\cdot
\nonumber
\\
& & 
\!\!\!\!\!\!\!\!\!\!\!\!\!\!\!
\cdot
\left(
a_l\overline{a_k}\hat{v}(-\sigma k\!+\!\tau l)+
a_l\overline{b_k}\hat{v}(-\!\sigma k\!-\!\tau l)+
b_l\overline{a_k}\hat{v}(\sigma k\!+\!\tau l)+
b_l\overline{b_k}\hat{v}(\!\sigma k\!-\!\tau l)
\right)
\mbox{ . }
\nonumber
\end{eqnarray}

\noindent Now $\EE\,v(l)^2=1$ implies that
$\EE\,\hat{v}(q)\hat{v}(p)=\frac{1}{L}\,\delta_{q,-p}$. 
With a bit of care, one can now check that the expectation value of
the product of the last two factors is equal to 1.

\vspace{.1cm}

{\rm (iii)} This is proven in the same manner as {\rm (ii)}.

\vspace{.1cm}

{\rm (iv)}
As $g_l=g_k=1$,
it follows from (\ref{eq-perturbcalc}) that
$W^*\tilde{P}W=W^*(P+P^*)W$ has only off-diagonal entries in
$(\pi_l+\pi_k)\CC^{2L}$ equal to
twice those of $W^*PW$. The first equality now follows directly from 
{\rm (ii)}, while the second one is checked similarly.

\vspace{.1cm}

{\rm (v)} One first verifies as above that (also for $l=k$ and
$\sigma=\pm$) 

$$
\EE\,\langle w_l^\sigma|P^*|\rw_k^\pm\rangle
\langle\rw_k^\pm| P|w_l^\sigma\rangle
\;=\;
\frac{1}{4L}\;h_k^2
\,h_l^2
\mbox{ . }
$$

\noindent Hence the claim follows by inserting
an identity
(\ref{eq-decomp}) and summing over $k$.
\hfill $\Box$

\section{Random dynamics of symplectic frames}
\label{sec-frames}

\subsection{Symplectic frames and isotropic manifolds}
\label{sec-unitaries}

The space $\Ff_p$ of symplectic $p$-frames, $p=1,\ldots,L$, is defined
by 

$$
\Ff_p
\;=\;
\left\{
(u_1,\ldots,u_p)\,\left|\,
u_l\in\RR^{2L}
\mbox{ , }
\langle u_l|u_k\rangle=\delta_{l,k}
\mbox{ , }
\langle u_l|J|u_k\rangle=0
\mbox{ , }
l,k=1,\ldots,p
\,
\right.\right\}
\mbox{ . }
$$

\noindent It is a manifold of dimension $p(2L-p)$. One could also call
$\Ff_p$ an isotropic Stiefel manifold and $\Ff_L$ the Langrangian
Stiefel manifold. 

\begin{proposi}
\label{prop-unitary}
The map $\zeta: \Ff_L\to\; ${\rm O}$(2L,\RR)\,\cap\,${\rm
SP}$(2L,\RR)\cong\,${\rm U}$(L,\CC)$
defined by

$$
\zeta(u)\;=\;(u,Ju)
\mbox{ , }
\qquad
u\;=\;(u_1,\ldots,u_L)\in\Ff_L
\mbox{ , }
$$

\noindent is an isomorphism.
\end{proposi}

\noindent {\bf Proof.} This is immediate if one recalls

$$
\mbox{O}(2L,\RR)\cap\mbox{SP}(2L,\RR)
\;=\;
\left\{
\left.
\left(
\begin{array}{cc}
a & b \\
-b & a \\
\end{array}
\right)
\;\right|
a,b\in \mbox{M}_{L\times L}(\RR)\mbox{ , }
a^ta+b^tb={\bf 1}
\mbox{ , }
a^tb=b^ta\;
\right\}
\mbox{ . }
$$

\noindent Moreover, $a+\imath\, b\in\,$U$(L,\CC)$ gives the second
isomorphism.
\hfill $\Box$

\vspace{.2cm}

A subspace $\Ee\subset\RR^{2L}$ is called
symplectic if $\langle v|J|v'\rangle=0$ for all
$v,v'\in\Ee$. The isotropic
manifold $L_p$ is by definition the set of all
oriented symplectic $p$-dimensional planes in $\RR^{2L}$.
It is a manifold of dimension $2p(L-p)+\frac{1}{2}\,p(p+1)$.
The maximal isotropic manifold $L_L$ is also called Lagrangian
manifold. 

\vspace{.2cm}

Next let $\Lambda^p\CC^{2L}$, $p=1,\ldots,L$, be the vector spaces 
of the anti-symmetrized
$p$-fold tensor products of $\CC^{2L}$. Decomposable (unentangled) vectors 
therein will be denoted
by $u_1\wedge\ldots\wedge u_p$ where $u_l\,\in\CC^{2L}$. 
A scalar product on $\Lambda^p\CC^{2L}$ is defined as usual by

$$
\langle
u_1\wedge\ldots\wedge u_p\,|\,
u'_1\wedge\ldots\wedge u'_p\rangle_{\Lambda^p\CC^{2L}}
\;=\;
\mbox{det}_p\left(\langle u_l\,|\,u'_k\rangle_{1\leq l,k\leq p}\right)
\mbox{ . }
$$

\noindent As is well-known, oriented $p$-dimensional
planes are isomorphic to the set of
decomposable real unit vectors in $\Lambda^p\CC^{2L}$. 
For the isotropic manifolds, this implies

$$
L_p
\;\cong\;
\left\{\left.
v_1\wedge\ldots\wedge v_p\in\Lambda^p
\RR^{2L}
\,\right|
\,
\|v_1\wedge\ldots\wedge v_p\|=1\;,\;
\langle v_l|J|v_k\rangle=0 \mbox{ , } l,k=1,\ldots,p
\right\}
\mbox{ . }
$$

Now each element $u=(u_1,\ldots,u_p)\in\Ff_p$ defines a sequence 
of embedded, oriented symplectic planes
$u_1\wedge\ldots\wedge u_q\in L_q$, $q=1,\ldots,p$.
Expressed in a different way, $u\in\Ff_p$ gives an element 
$u_1\wedge\ldots\wedge u_p\in L_p$ as well as an unoriented, but
ordered  
orthonormal basis therein. This is locally an isomorphism:

\begin{proposi}
$\Ff_p$ is a principal bundle over $ L_p$ with fiber {\rm
SO}$(p,\RR)$. 
\end{proposi}


The following elementary
lemma about matrix elements of a Lagrangian projection
w.r.t.  eigenvectors of $J$ (and hence also all rotations $R$ constructed in
Section \ref{sec-normal})
will be used later on.

\begin{lemma}
\label{lem-lagrangian}
Let $\Pi=uu^t$ be the projection on the Lagrangian plane associated with
$u\in\Ff_L$. If $v_j\in\CC^{2L}$, $j=1,2$, are two 
normalized orthogonal eigenvectors of $J$,
namely $Jv_j=\imath \sigma_j v_j$ for signs $\sigma_j$,
then

$$
(1+\sigma_j\sigma_k)\;\langle v_j|\Pi|v_k\rangle
\;=\;
\delta_{j,k}
\mbox{ . }
$$

\end{lemma}

\noindent {\bf Proof.} Let $\Pi_1$ and $\Pi_2$ be the projections on
the first and second $L$ components of $\CC^{2L}$ respectively.
Hence  $\Pi_1+\Pi_2={\bf 1}$.
Using the orthogonal $(u,Ju)$ of Proposition \ref{prop-unitary}, one
then has

$$
\delta_{j,k}
\;=\;
\langle v_j|(u,Ju)(\Pi_1+\Pi_2)
(u,Ju)^t
|v_k\rangle
\;=\;
\langle v_j|uu^t
|v_k\rangle
+
\langle v_j|Juu^tJ^t
|v_k\rangle
\mbox{ . }
$$

\noindent As $J^t=J^*=-J$, the claim follows from the supposed 
properties of $v_j$.
\hfill $\Box$

\subsection{Action of a transfer matrix on a symplectic frame}
\label{sec-action}

The group SP$(2L,\RR)$ acts on the space $\Ff_L$ of symplectic
$L$-frames. 
There is an obvious way to define such an action
$\Uu:\,$SP$(2L,\RR)\times\Ff_L\to\Ff_L$
(which will later on actually turn out 
to be relevant for the calculation of the Lyapunov exponents):
given $T\in\,$SP$(2L,\RR)$ and $u=(u_1,\ldots,u_L)\in \Ff_L$,
the plane $Tu_1\wedge\ldots\wedge T u_p$ is symplectic for any $p\leq L$; 
hence applying the Schmidt orthonormalization procedure to the sequence
$Tu_1,\ldots,Tu_L$ gives a new element $\Uu_T u \in\Ff_L$. On a
calculatory level, it will be convenient (and equivalent as one easily
verifies) to define this action
using wedge products:

\begin{equation}
\label{eq-action}
\Uu_T \;u_1\wedge\ldots\wedge u_p
\;=\;
\frac{\Lambda^p T\,u_1\wedge\ldots\wedge u_p}{
\|\Lambda^p T\,u_1\wedge\ldots\wedge u_p\|}
\mbox{ , }
\qquad
p=1,\ldots,L
\mbox{ . }
\end{equation}

\noindent More explicitly, this means that the $p$th vector of the new
frame $(\Uu_T \;u)_p$ satisfies for all
$v\in\CC^{2L}$:

\begin{equation}
\label{eq-action2}
\langle v|(\Uu_T \;u)_p\rangle
\;=\;
\frac{\langle (\Lambda^{p-1} T\,u_1\wedge\ldots\wedge u_{p-1})\wedge v
|\Lambda^{p} T\,u_1\wedge\ldots\wedge u_{p}\rangle}{
\|\Lambda^{p-1} T\,u_1\wedge\ldots\wedge u_{p-1}\|
\;\|\Lambda^p T\,u_1\wedge\ldots\wedge u_p\|}
\mbox{ . }
\end{equation}

\noindent
It is immediate from the definition that 

$$
\Uu_{ST}\;=\;\Uu_S\,\Uu_T
\mbox{ , }
\qquad 
S,T\in\,\mbox{SP}(2L,\RR)
\mbox{ . }
$$

For fixed $p$, (\ref{eq-action}) defines a map $\Uu_T$ on $L_p$. But the
whole sequence $p=1,\ldots,L$, defines a map on $\Ff_L$. 
Due to Proposition \ref{prop-unitary}, this defines an action
on the unitary group U$(L,\CC)$. 
Therefore each $\Uu_T$ can be identified with
an element of U$(L,\CC)$ itself (explaining hence the notation with a
letter $\Uu$).

\subsection{Definition of random dynamics}
\label{sec-dynamics}

The random transfer matrices are symplectic so that they induce an
action on the frames.
Here their transformed normal form 
$R(1-\lambda P(n))$ given by
(\ref{eq-normal}) will be used to
define a random dynamical system on $\Ff_L$.
The random orbits in $\Ff_L$ for some fixed initial condition
$(u_1(0),\ldots,u_L(0))$
will be defined and
denoted as follows:

\begin{equation}
\label{eq-randact}
u_1(n)\wedge\ldots\wedge u_L(n)
\;=\;
\Uu_{R(1-\lambda P(n))}
u_1(n-1)\wedge \ldots\wedge u_L(n-1)
\mbox{ , }
\qquad
n\geq 1
\mbox{ . }
\end{equation}

The free dynamics at $\lambda=0$ is non-random and just given by the
(elliptic-hyperbolic) rotation $R$, the analysis of which is
straight-forward. The main objects of this work is to study the
effect of weakly coupled randomness.

\vspace{.2cm}

All the information needed for calculating averaged quantities like
the Lyapunov exponent can actually be encoded in an invariant
measure $\nu$ on $\Ff_L$. It can be defined by

$$
\int_{\Ff_L} d\nu(u)\;
f(u)\;=\;
\EE\,\int_{\Ff_L} d\nu(u)\;
f(\Uu_{R(1-\lambda P)} u)
\mbox{ . }
$$

\noindent It is known \cite{BL} that $\nu$ is unique as soon as
$\lambda\neq 0$ and moreover
continuous. In \cite[Section 4.7]{JSS}, 
perturbation theory for this measure
(around the Lebesgue measure) was done in the case $L=1$.
The results below (in particular, Section \ref{sec-osci}) can be
interpreted in a similar way.

\subsection{Channel weights}
\label{sec-weights}

It will be useful to introduce the probability (or weight) 
$\rho_{p,j}(n)\in[0,1]$ for the $p$th frame vector $u_p(n)$ to be in the
$j$th channel. More explicitly,

\begin{equation}
\rho_{p,j}^\pm(n)
\;=\;
\langle u_p(n)|\pi^\pm_j|u_p(n)\rangle
\mbox{ , }
\qquad
\rho_{p,j}(n)\;=\;\rho_{p,j}^+(n)+\rho_{p,j}^-(n)
\label{eq-defweights}
\mbox{ . }
\end{equation}

\noindent 
The identity (\ref{eq-decomp}) and Lemma \ref{lem-lagrangian} (use $2\nu_j$
eigenvectors of $J$ to build a basis in $\pi_j\CC^{2L}$) imply
respectively 

\begin{equation}
\label{eq-magicid}
\sum_{j=0}^{L_c}\rho_{p,j}(n)\;=\;1
\mbox{ , }
\qquad
\sum_{p=1}^{L}\rho_{p,j}(n)\;=\;
\nu_j
\mbox{ . }
\end{equation}

\noindent 
For an elliptic channel one has $\pi_j^-=\overline{\pi_j^+}$ so that

\begin{equation}
\label{eq-ellipweight}
\rho_{p,j}(n)
\;=\;
2\;\rho_{p,j}^\pm(n)
\mbox{ , }
\qquad
\mu_j\;\mbox{ elliptic .}
\end{equation}

\noindent For the random dynamics of frames, let us introduce the 
mean presence probability of the $p$th frame vector in the $j$th channel:

\begin{equation}
\label{eq-meanweights}
\langle \rho_{p,j}\rangle_N
\;=\;
\frac{1}{N}\sum_{n=0}^{N-1}\;
\EE\,\rho_{p,j}(n)
\mbox{ , }
\qquad
\langle \rho_{p,j}\rho_{q,k}\rangle_N
\;=\;
\frac{1}{N}\sum_{n=0}^{N-1}\;
\EE\,\rho_{p,j}(n)\rho_{q,k}(n)
\mbox{ . }
\end{equation}

\noindent Similarly, higher moments are defined. 
It follows from a standard ergodic argument that these quantities
converge in the limit $N\to\infty$ to some numbers denoted
$\langle \rho_{p,j}\rangle$, $\langle\rho_{p,j}\rho_{q,k}\rangle$ and
so on. However, this fact will not be
used below. 

\subsection{Separating hyperbolic from elliptic channels}
\label{sec-separ}

A bit of thought shows that the first $2L_h+1$ frame vectors
$u_1,\ldots,u_{2L_h+1}$
deterministically aline (up to an error) 
with the expanding 
hyperbolic basis vectors, that is $\rw^+_1,\ldots,\rw^+_{L_h},
\rw^+_{L-L_h},\ldots,\rw^+_{L}$ (recall that we supposed
$\mu_l<2$, as in Fig.~1(i)). 
Hence the remaining frame vectors have to be in
the elliptic channels due to orthogonal and symplectic blocking. Hence
let us call the frame vectors $u_{2L_h+2},\ldots,u_L$ elliptic, while
the first ones are called hyperbolic.
The corresponding analysis, elementary but 
a bit tedious, is carried out below. It is
not needed in case all channels are elliptic (as in Fig.~1(ii)). 

\begin{proposi}
\label{prop-separ}
For almost
every initial condition {\rm (}or almost every disorder
configuration{\rm )}, 
for $n$ sufficiently large and for $k=1,\ldots,L_h$,

$$
|\langle \rw^+_L|u_1(n)\rangle|^2
\;=\;
1-\Oo(\lambda^2)
\mbox{ , }
\qquad
|\langle \rw^+_{k}\wedge \rw^+_{L-k}|u_{2k}(n)\wedge u_{2k+1}(n)\rangle|^2
\;=\;
1-\Oo(\lambda^2)
\mbox{ . }
$$

\end{proposi}

\vspace{.2cm}

\noindent {\bf Proof.} 
Let us begin with the study of $|\langle \rw^+_L|u_1(n)\rangle|^2$. 
It follows from (\ref{eq-action}) and (\ref{eq-randact}) that

\begin{equation}
|\langle \rw^+_L|u_1(n+1)\rangle|^2
\;=\;
\frac{e^{2\eta_L}\,|\langle \rw^+_L|u_1(n)\rangle|^2}{
\langle u_1(n)|R^*R |u_1(n)\rangle}\;+\;\Oo(\lambda)
\mbox{ . }
\label{eq-historyhyp}
\end{equation}

\noindent In order to analyse the denominator, let us
insert (\ref{eq-decomp}):

\begin{eqnarray}
\langle u_1(n)|R^*R |u_1(n)\rangle
& = &
\sum_{l=1}^L\,e^{\eta_l(1-g_l^2)}
|\langle \rw^+_l|u_1(n)\rangle|^2
+e^{-\eta_l(1-g_l^2)}
|\langle \rw^-_l|u_1(n)\rangle|^2
\nonumber
\\
& = &
1+
\sum_{l=0}^{L_h}\,(e^{2\eta_l}-1)\,
|\langle \rw^+_l|u_1(n)\rangle|^2
+
(e^{-2\eta_l}-1)
|\langle \rw^-_l|u_1(n)\rangle|^2
\mbox{ , }
\nonumber
\end{eqnarray}

\noindent where as before we identified indices $L\widehat{=}0$ and
used that $u_1(n)$ is normalized.
The next aim is to prove an upper bound on this and therefore let us
first note that $e^{-2\eta_l}-1<0$ so that 
those terms can be discarded. 
For the remainder, starting from the smallest factor 
$e^{2\eta_l}-1$ and going iteratively to the largest, one gets
using each time $|\langle \rw^+_l|u_1(n)\rangle|^2\leq 1-
\sum_{k=0}^{l-1}|\langle \rw^+_k|u_1(n)\rangle|^2$,

\begin{eqnarray}
\langle u_1(n)|R^*R |u_1(n)\rangle
& \leq &
e^{2\eta_{L_h}}+
\sum_{l=0}^{L_h-1}
\,(e^{2\eta_l}-e^{2\eta_{L_h}})\,
|\langle \rw^+_l|u_1(n)\rangle|^2
\nonumber
\\
\label{eq-denomest}
& &
\\
& \leq &
e^{2\eta_1}+
(e^{2\eta_L}-e^{2\eta_1})\,
|\langle \rw^+_L|u_1(n)\rangle|^2
\mbox{ . }
\nonumber
\end{eqnarray}

\noindent Note that $\eta_1$ is the second largest hyperbolic
angle satisfying $\eta_1<\eta_L$. Replacing in the above, 

$$
|\langle \rw^+_L|u_1(n+1)\rangle|^2
\;\geq\;
\frac{e^{2(\eta_L-\eta_1)}\,|\langle \rw^+_L|u_1(n)\rangle|^2}{
1+(e^{2(\eta_L-\eta_1)}-1)\,|\langle \rw^+_L|u_1(n)\rangle|^2
}\;+\;\Oo(\lambda)
\mbox{ . }
$$

\noindent Thus, up to an error, 
$|\langle \rw^+_L|u_1(n+1)\rangle|^2$ is larger than the image of
$x=|\langle \rw^+_L|u_1(n)\rangle|^2$ under the function
$f_a(x)=\frac{ax}{1+(a-1)x}$ where
$a=e^{2(\eta_L-\eta_1)}>1$. As $n$ grows, this procedure is then
iterated, giving rise to a discrete time dynamics through successive
application of $f_a$. The
function $f_a$ has two fixed points in $[0,1]$, an unstable one at $0$
and a stable one at $1$. Either the initial condition is already away
from the unstable fixed point or the random perturbation leads the
discrete time dynamics to leave it (only with exponentially small
probability one remains in its neighborhood, as an elementary argument
shows). As it only takes a finite number of iterations 
to get within the neighborhood
of the stable fixed point and the (random) perturbation is of order
$\Oo(\lambda)$,  one can conclude that, for $n$ large enough,

$$
|\langle \rw^+_L|u_1(n)\rangle|^2
\;=\;1-\Oo(\lambda)
\mbox{ . }
$$

\noindent When this holds, however, 
the random perturbation cannot be linear in
$\lambda$ anymore, because the $ \Oo(\lambda)$-term
would not have definite sign and hence violate
$|\langle \rw^+_L|u_1(n)\rangle|^2\leq 1$. Indeed, it is elementary to
verify also algebraically that the perturbative terms linear in
$\lambda$ vanish in (\ref{eq-historyhyp}) when one already knows
$\langle \rw^+_L|u_1(n)\rangle=1-\Oo(\lambda)$. Hence one can repeat the
above argument in the neighborhood of the stable fixed point $1$, but
based on (\ref{eq-historyhyp}) with an error term $\Oo(\lambda^2)$.
This implies the
first claim. Moreover, due to normalization, $|\langle
{\rw}^+_l|u_1(n)\rangle|^2=\Oo(\lambda^2)$ for all $l\neq L$ as well as
$|\langle
\rw^-_l|u_1(n)\rangle|^2=\Oo(\lambda^2)$ for any $l$.

\vspace{.2cm}

The remaining estimates are proven by recurrence over $k$. After
having exploited orthogonal and
symplecting blocking, the basic argument is as before
and therefore some calculatory 
details are suppressed and left to the reader.  Hence let
us suppose that $|\langle \rw^+_{l}\wedge 
\rw^+_{L-l}|u_{2l}(n)\wedge u_{2l+1}(n)\rangle|^2=
1-\Oo(\lambda^2)$ for all $l<k$. As $\langle
u_{2k}(n)|u_m(n)\rangle=\delta_{2k,m}$ and $\langle
u_{2k}(n)|J|u_m(n)\rangle=0$, one concludes that
$\langle
u_{2k}(n)|\rw^+_m\rangle=\Oo(\lambda)$ and $\langle
u_{2k}(n)|J|\rw^+_m\rangle=\Oo(\lambda)$ for 
$m=1,\ldots,k-1,L-k+1,\ldots L$. But for these hyperbolic channels 
$J|\rw^+_m\rangle=|\rw^-_m\rangle$. As the same holds for
$u_{2k+1}(n)$, we can conclude that 

$$
\langle \rw_m^\sigma\wedge\rw_l^\tau |u_{2k}(n)\wedge
 u_{2k+1}(n)\rangle
\;=\;\Oo(\lambda)
\mbox{ , }
\qquad
l,m=1,\ldots,k-1,L-k+1,\ldots L
\mbox{ . }
$$

\noindent Using this, a short perturbative calculation
starting from (\ref{eq-action2}) shows

$$
|\langle \rw^+_k\wedge\rw^+_{L-k} |u_{2k}(n+1)\wedge u_{2k+1}(n+1)\rangle|^2
\,=\,
\frac{e^{4\eta_k}\;
|\langle \rw^+_k\wedge\rw^+_{L-k} |u_{2k}(n)\wedge u_{2k+1}(n)\rangle|^2}{
\langle u_{2k}(n)\wedge u_{2k+1}(n)|\Lambda^2R^*R
|u_{2k}(n)\wedge u_{2k+1}(n)\rangle
}+\Oo(\lambda).
$$

%

\noindent Generalizing the argument leading to
(\ref{eq-denomest}), one can bound the
denominator from above by

$$
e^{4\eta_{k+1}}\,+\,(e^{4\eta_k}-e^{4\eta_{k+1}})\,
|\langle \rw^+_k\wedge\rw^+_{L-k} |u_{2k}(n)\wedge u_{2k+1}(n)\rangle|^2
\;+\Oo(\lambda^2)
\;.
$$

\noindent As $\eta_k>\eta_{k+1}$, one can use the same function $f_a$
as above with $a=e^{4(\eta_k-\eta_{k+1})}$ and complete the
(two-stepped) argument
as above.
\hfill $\Box$

\vspace{.2cm}

\section{Lyapunov exponents}
\label{chap-lyap}

\subsection{Calculating Lyapunov exponents with symplectic frames}
\label{sec-lyapinit}

In the definition (\ref{eq-lyapform}) of the Lyapunov exponents
appears the operator norm. Instead, one may use 
symplectic planes as initial condition if an averaging over them is
done. This is briefly discussed in this section.

\vspace{.2cm}

Important is the
well-known fact \cite{BL} that for any symplectic matrix $T$, its
second quantized is most
expansive on the  isotropic subspaces, namely the norm of
the second quantized $\Lambda^p T$ can be calculated by
$\|\Lambda^p T\|=\sup_{u\in L_p}\|\Lambda^p T u\|=
\sup_{u\in \Ff_p}\|\Lambda^p T u\|$.
Furthermore, the Lyapunov exponents
according to \cite[A.III.3.4]{BL} the Lyapunov exponents are given by 

\begin{equation}
\label{eq-lyapform}
\sum_{l=1}^p\,\gamma_l
\;=\;
\lim_{N\to\infty}\,\frac{1}{N}\;
\EE\,\log
\left(\,\left\|\;\prod_{n=1}^N\Lambda^pT(n)\;u(0)
\;\right\|_{\Lambda^p\CC^{2L}}\right)
\mbox{ , }
\end{equation}

\noindent where  $u(0)\in\Ff_p$ is an arbitrary initial condition. 
One may average over $u(0)$ w.r.t. to the invariant measure 
$\nu$ on $\Ff_L$ and this immediately leads to

$$
\sum_{l=1}^p\,\gamma_l
\;=\;
\int_{\Ff_p}d{\nu}(u)\;
\EE\,\log
\left(\,\left\|\;\Lambda^pT\;u
\;\right\|_{\Lambda^p\CC^{2L}}\right)
\mbox{ , }
$$

\noindent where here the $\EE$ is only an average over the single site
transfer matrix $T$. Similar formulas can be found in 
\cite[Section IV.6]{CL}.

\subsection{Basic perturbative formula: only symplectic channels}
\label{sec-lyapsymp}

As it is considerably more transparent, let us first perform the
perturbative calculation of the Lyapunov exponents in the
case where there are no hyperbolic channels (as in Fig.~1(ii)). 
Hence we assume $R$ to be orthogonal.

\vspace{.2cm}

Let us insert ${\bf 1}=\Lambda^pM\Lambda^p M^{-1}$ 
in between each pair of transfer matrices
in (\ref{eq-lyapform}) (the boundary terms do not change anything as
can easily be argued as in \cite[Section 4.1]{JSS}, for example). 
Then develop the product therein into a
telescopic sum using the definition of the action (\ref{eq-action}) as
well as the definition of the random dynamics of frames. This gives

$$
\sum_{l=1}^p\,\gamma_l
\;=\;
\lim_{N\to\infty}\,\frac{1}{N}\;
\sum_{n=0}^{N-1}\;
\EE\,\log
\left(\left\|
\Lambda^p (M^{-1}T(n+1)M)\;
u_1(n)\wedge\ldots\wedge
u_p(n)\right\|_{\Lambda^p\CC^{2L}}
\right)
\mbox{ , }
$$

\noindent where $\EE$ contains also an average over the initial
condition $u_1(0)\wedge\ldots\wedge
u_p(0)$. As $R$ is orthogonal, so is $\Lambda^p R$. Using
$\Lambda^p (M^{-1}T(n)M)=\Lambda^pR 
\;\Lambda^p(1-\lambda P(n))$, 
one gets writing out the norm explicitly:

$$
\sum_{l=1}^p\,\gamma_l
\;=\;
\frac{1}{2}\;
\lim_{N\to\infty}\,\frac{1}{N}\;
\sum_{n=0}^{N-1}\;
\EE\,\log
\left(
\mbox{det}_p
\left(\langle (1-\lambda P(n+1))u_l(n)\,|\,
(1-\lambda P(n+1))
u_k(n)\rangle_{1\leq l,k\leq p}\right)
\right)
\mbox{ .}
$$

\noindent Now $\log \mbox{det}_p=\mbox{Tr}_p\log$, so that multiplying
out gives:

$$
\sum_{l=1}^p\,\gamma_l
\;=\;
\frac{1}{2}\;
\lim_{N\to\infty}\,\frac{1}{N}\;
\sum_{n=0}^{N-1}\;
\EE\,
\mbox{Tr}_p\log
\left(
{\bf 1}_p+
\langle u_l(n)\,|(-\lambda (P+P^*)+\lambda^2 |P|^2)|\,
u_k(n)\rangle_{1\leq l,k\leq p}
\right)
\mbox{ , }
$$

\noindent where the argument $n+1$ of $P=P(n+1)$ was suppressed because they
are all independent and identically distributed random variables over
each of which can be averaged independently in each summand. Finally
let $\Pi_p(n)$ be the projection in $\RR^{2L}$ onto the subspace
spanned by $u_1(n),\ldots,u_p(n)$. Expanding the logarithm up to order
$\Oo(\lambda^3)$ and using
that $\EE\, \mbox{Tr}(\Pi_p(n)P)=0$, we obtain

\begin{equation}
\label{eq-lyapexpan0}
\sum_{l=1}^p\,\gamma_l
\;=\;
\frac{\lambda^2}{2}\,
\lim_{N\to\infty}\,\frac{1}{N}\;
\sum_{n=0}^{N-1}\;
\EE
\left(
\mbox{Tr}(\Pi_p(n)|P|^2)
-\frac{1}{2}\,
\mbox{Tr}(\tilde{P}\Pi_p(n)\tilde{P}\Pi_p(n))
\right)
+\Oo(\lambda^3)\;,
\end{equation}

\noindent where the trace is now over $\RR^{2L}$ and $\tilde{P}
=P+P^*$ is a real and self-adjoint matrix.
Subtracting gives, up to $\Oo(\lambda^3)$,

\begin{equation}
\label{eq-lyapexpan}
\gamma_p=
\frac{\lambda^2}{2}
\lim_{N\to\infty}\,\frac{1}{N}
\sum_{n=0}^{N-1}
\EE
\left(
\langle u_p(n)||P|^2|u_p(n)\rangle
-
\langle u_p(n)|\tilde{P}\Pi_{p}(n)\tilde{P}|u_p(n)\rangle
+
\frac{1}{2}
\langle u_p(n)|\tilde{P}|u_p(n)\rangle^2
\right)
.
\end{equation}

\noindent As we shall show in the next section, the first and second
contribution cancel exactly for the bottom Lyapunov exponent,  
while the third one can be calculated
explicitly, namely it follows directly from Lemma \ref{lem-osci} below
(set $L_h=-1$ therein so that there are no hyperbolic channels) that

\begin{equation}
\label{eq-lyapexpansymp}
\gamma_L
\;= \;
\lim_{N\to\infty}\;
\frac{\lambda^2}{8L}
\sum_{j,k=0}^{L_c}h_j^2h_k^2
(2-\delta_{j,k})\;
\langle\rho_{L,j}\rho_{L,k}\rangle_N
+\Oo(\lambda^3)
\,.
\end{equation}

\noindent Because $h_j^2\geq 1$, Theorem
\ref{theo-main} follows immediately in the case where there are only
elliptic channels.

\vspace{.2cm}

\subsection{Oscillatory sums}
\label{sec-osci}

The aim of this section is to evaluate the terms appearing in the
perturbative expansion (\ref{eq-lyapexpan}) of the Lyapunov
exponent. This will then give (\ref{eq-lyapexpansymp}). 
As it does not take more effort at this point and will be needed
below, we will however not suppose all channels to be elliptic, but
only the exterior frame vector $u_p$ to be elliptic,
{\sl i.e.} $p>2L_h+1$.

\begin{lemma} 
\label{lem-osci} Let $p>2L_h+1$ and suppose that the Main hypothesis
holds. 

\vspace{.1cm}

\noindent {\rm (i)}  
$$
\frac{1}{N}\sum_{n=0}^{N-1}\;
\EE
\,
\langle u_p(n)|\,|RP|^2\,|u_p(n)\rangle
\;=\;
\frac{1}{2}\;h_\av^2\,
\sum_{l=0}^{L_c}
h_l^2
\;\langle
\rho_{p,l}
\rangle_N
\;+\;\Oo(N^{-1},\lambda)
\;.
$$

\noindent {\rm (ii)} 
$$
\frac{1}{N}\sum_{n=0}^{N-1}\;
\EE
\,
\langle u_p(n)|\tilde{P}|u_p(n)\rangle^2
\;=\;
\frac{1}{2L}
\sum_{j,k>L_h}^{L_c}h_j^2h_k^2(2-\delta_{j,k})\,
\;\langle
\rho_{p,j}\rho_{p,k}
\rangle_N
\;+\;\Oo(N^{-1},\lambda)
\;.
$$

\noindent {\rm (iii)} 
$$
\frac{1}{N}\sum_{n=0}^{N-1}\;
\EE
\,
\sum_{q=2L_h+2}^L\langle u_p(n)|\tilde{P}
|u_q(n)\rangle^2
\;=\;
\frac{1}{2}\,\left(\frac{1}{L}
\sum_{l>L_h}^{L_c}\nu_lh_l^2\right)
\;
\sum_{k=0}^{L_c}
h_k^2\;
\langle
\rho_{p,k}
\rangle_N
\;+\;\Oo(N^{-1},\lambda)
\;.
$$

\end{lemma}

\noindent {\bf Proof.} {\rm (i)}
By inserting identities (\ref{eq-decomp}),

\begin{equation}
\label{eq-help2}
\langle u_p(n)|\,|RP|^2\,|u_p(n)\rangle
\;=\;
\sum_{l,k=0}^{L_c}\;
\langle u_p(n)|(\pi_l^++\pi_l^-)\,|RP|^2\,(\pi_k^++\pi_k^-)|u_p(n)\rangle
\mbox{ . }
\end{equation}

\noindent By Proposition \ref{prop-separ}, the sum may be restricted
to $L_h+1\leq l,k\leq L_c$ at the cost of an error $\Oo(\lambda)$.
Hence let us consider, for fixed elliptic channels $l,k$, and signs
$\sigma,\tau$, 

$$
J(N)\;=\;
\frac{1}{N}\sum_{n=0}^{N-1}\;
\EE\;
\langle u_p(n)|\pi_l^\sigma\,|RP|^2\,\pi_k^\tau|u_p(n)\rangle
\mbox{ . }
$$
\noindent As from (\ref{eq-action}),
$$
\pi_k^\tau|u_p(n)\rangle
\;=\;
e^{\tau\imath \eta_k}
\pi_k^\tau|u_p(n-1)\rangle\;+\;\Oo(\lambda)
\mbox{ , }
$$
\noindent we get, because the boundary terms are
of $\Oo(N^{-1})$,

$$
J(N)\;=\;e^{\imath(-\sigma\eta_l+\tau\eta_k)}\,J(N)
\;+\;\Oo(N^{-1},\lambda)
\mbox{ . }
$$

\noindent If now $e^{\imath(-\sigma\eta_l+\tau\eta_k)}\neq 1$, this implies
$J(N)=\Oo(N^{-1},\lambda)$. By the main hypothesis this does not happen if
$l\neq k$ or $\sigma\neq\mu$. 
Therefore only the diagonal terms in
(\ref{eq-help2}) contribute to leading order so that

$$
\frac{1}{N}\sum_{n=0}^{N-1}\;
\EE\,\langle u_p(n)|\,|RP|^2\,|u_p(n)\rangle
\;=\;
\sum_{l=L_h+1}^{L_c}
\frac{1}{N}\sum_{n=0}^{N-1}\;
\sum_{\sigma=\pm}\,\EE\,
\langle u_p(n)|\pi_l^\sigma|RP|^2\pi_l^\sigma|u_p(n)\rangle
\;+\;\Oo(N^{-1},\lambda)
\,.
$$

\noindent Finally $\pi_l^\sigma|u_p(n)\rangle=
(\frac{1}{2}\,\rho_{p,l}(n))^{\frac{1}{2}}
\,|w_l^\sigma\rangle$ for some complex unit 
vector $w_l^\sigma$
satisfying $\pi_l^\sigma|w_l^\sigma\rangle=|w_l^\sigma\rangle$. Thus

$$
\frac{1}{N}\sum_{n=0}^{N-1}\;
\EE\,\langle u_p(n)|\,|RP|^2\,|u_p(n)\rangle
\;=\;
\frac{1}{N}\sum_{n=0}^{N-1}\;
\sum_{l=L_h+1}^{L_c}
\,\sum_{\sigma=\pm}
\;\EE\,
\frac{1}{2}\,\rho_{p,l}(n)\;
\langle w_l^\sigma|\,|RP|^2\,|w_l^\sigma\rangle
+\Oo(N^{-1},\lambda)
\,.
$$

\noindent But the expectation value of the matrix element (over the
random variable $P$ only) is independent of $w_l^\sigma$ and given by Lemma
\ref{lem-matrixel}(v). This directly leads to the first claim
because the sum can again be extended to $l=0,\ldots,L_c$ by
Proposition \ref{prop-separ}.

\vspace{.2cm}

{\rm (ii)} One has for $p,q>2L_h+1$
$$
\langle u_p(n)|\tilde{P}|u_q(n)\rangle^2
\;=\;
\sum_{k,l,m,j=0}^{L_c}\;
\sum_{\sigma_k,\sigma_l,\sigma_m,\sigma_j=\,\pm}
\;
\langle u_p(n)|\pi_k^{\sigma_k}\tilde{P}\pi_l^{\sigma_l}|u_q(n)\rangle
\;
\langle u_q(n)|\pi_m^{\sigma_m}\tilde{P}\pi_j^{\sigma_j}|u_p(n)\rangle
\mbox{ . }
$$

\noindent For the same reason as above, the sum can be restricted to
elliptic channels $k,l,m,j> L_h$ up to errors of order $\Oo(\lambda)$.
From the 16 signs, Lemma \ref{lem-matrixel}(iv) eliminates
half, forcing $\sigma_k=-\sigma_l$ and $\sigma_m=-\sigma_j$.
To each of the finite number of remaining summands an oscillatory sum
argument will now be applied. Set

$$
J(N)\;=\;
\frac{1}{N}\sum_{n=0}^{N-1}\;
\EE\;
\langle u_p(n)|\pi_k^{-\sigma}\tilde{P}\pi_l^{\sigma}|u_q(n)\rangle
\;
\langle u_q(n)|\pi_m^{-\tau}\tilde{P}\pi_j^{\tau}|u_p(n)\rangle
\mbox{ . }
$$

\noindent Proceeding as above shows

$$
J(N)\;=\;
e^{\imath\sigma(\eta_k+\eta_l)}\,e^{\imath\tau(\eta_m+\eta_j)}
\,J(N)
\;+\;\Oo(N^{-1},\lambda)
\mbox{ . }
$$

\noindent Again invoking the main hypothesis,
$J(N)=\Oo(1)$ is therefore possible
only if $\tau=-\sigma$ and $\{k,l\}=\{m,j\}$.
Hence up to $\Oo(N^{-1},\lambda)$,

\begin{eqnarray}
& & \!\!\!\!\!\!\!\!\!\!\!\!\!\!\!
\frac{1}{N}\sum_{n=0}^{N-1}\; 
\EE\,\langle u_p(n)|\tilde{P}|u_q(n)\rangle^2
\nonumber 
\\
&  & =\;
\frac{1}{N}\sum_{n=0}^{N-1}\; 
\sum_{k,l>L_h}^{L_c} 
\,\sum_{\sigma=\pm} 
\;\EE\, 
\langle u_p(n)|\pi_k^{-\sigma}\tilde{P}\pi_l^{\sigma}|u_q(n)\rangle
\;
\langle u_q(n)|\pi_l^{\sigma}\tilde{P}\pi_k^{-\sigma}|u_p(n)\rangle
\label{eq-help3}
\\
&  & \;\;\;\;\; +\;
\frac{1}{N}\sum_{n=0}^{N-1}\; 
\sum_{k,l>L_h,\,k\neq l}^{L_c} 
\,\sum_{\sigma=\pm} 
\;\EE\, 
\langle u_p(n)|\pi_k^{-\sigma}\tilde{P}\pi_l^{\sigma}|u_q(n)\rangle
\;
\langle u_q(n)|\pi_k^{\sigma}\tilde{P}\pi_l^{-\sigma}|u_p(n)\rangle
\;.
\label{eq-help4}
\end{eqnarray}

\noindent Normalizing the projections of the frame vectors and then 
applying Lemma \ref{lem-matrixel}(iv) gives

$$
(\ref{eq-help3})
\;=\;
\frac{1}{2L}\;
\sum_{ k,l>L_h}^{L_c} 
h_l^2\,h_k^2\;
\left(
\frac{1}{N}\sum_{n=0}^{N-1}\; 
\EE\;\rho_{p,k}(n)\,\rho_{q,l}(n)
\right)
\mbox{ . }
$$

\noindent The contribution (\ref{eq-help4}) can only be treated
similarly if $q=p$. Supposing this, the second identity in 
Lemma \ref{lem-matrixel}(iv) shows

$$
(\ref{eq-help4})
\;=\;
\frac{1}{2L}\;
\sum_{k,l>L_h,\,k\neq l}^{L_c} 
h_l^2\,h_k^2\;
\left(
\frac{1}{N}\sum_{n=0}^{N-1}\; 
\EE\;\rho_{p,k}(n)\,\rho_{p,l}(n)
\right)
\mbox{ . }
$$

\noindent The sum of the latter two contributions is given in (ii).

\vspace{.2cm}

(iii) One now has to sum (\ref{eq-help3}) and 
(\ref{eq-help4}) over $q=2L_h+2,\ldots, L$,
namely precisely the elliptic frame vectors.
But because $k$ and $l$ only correspond to elliptic channels, the sum
may be extended to $q=1,\ldots,L$ because the weight of the hyperbolic
frame vectors in the elliptic channels is of order $\Oo(\lambda)$ by
Proposition \ref{prop-separ}. It now follows that the contribution of 
(\ref{eq-help4}) vanishes. In order to show this, 
decompose $\pi^\sigma_k$ and $\pi^\sigma_l$ therein using
(\ref{eq-basisdecom}) and note that the directions $|\rw_k^\sigma\rangle$ and 
$|\rw_l^\sigma\rangle$ on which they project satisfy the hypothesis of Lemma 
\ref{lem-lagrangian} due to the identities (\ref{eq-basisJ}).
Therefore

\begin{equation}
\label{eq-help5}
\sum_{q=1}^{L}\;
\pi_k^{\sigma}|u_q(n)\rangle
\;
\langle u_q(n)|\pi_l^{\sigma}
\;=\;
0
\mbox{ , }
\qquad
k\neq l
\mbox{ , }
\end{equation}

\noindent implying the claim. 
The sum of (\ref{eq-help3}) over $q=1,\ldots,L$ can easily be
carried out using the identity (\ref{eq-magicid}):

$$
\sum_{q=1}^L\;
(\ref{eq-help3})
\;=\;
\frac{1}{2L}\;
\sum_{ l>L_h}^{L_c} 
h_l^2\,\nu_l
\;
\sum_{ k>L_h}^{L_c} 
h_k^2\,
\left(
\frac{1}{N}\sum_{n=0}^{N-1}\; 
\EE\;\rho_{p,k}(n)\,
\right)
\mbox{ . }
$$

\noindent Because $u_p$ is elliptic, the sum may carry over
$k=0,\ldots,L_c$ because the error is $\Oo(\lambda^2)$.
\hfill $\Box$

%

\subsection{Sum of Lyapunov exponents near band center}
\label{sec-perturbativesum}

Let us again suppose in this section that there are only elliptic
channels. Then it follows from
(\ref{eq-lyapexpan0}) and Lemma \ref{lem-osci} that

\begin{eqnarray}
\sum_{l=1}^L\,\gamma_l
& = &
\frac{\lambda^2}{2}\,
\lim_{N\to\infty}\,\frac{1}{N}\;
\sum_{n=0}^{N-1}\,
\sum_{l=1}^L\,
\EE
\left(
\langle u_l(n)|\,|P|^2|u_l(n)\rangle
-\frac{1}{2}\,
\langle u_l(n)|\tilde{P}\Pi_L(n)\tilde{P}|u_l(n)\rangle
\right)
+\Oo(\lambda^3)
\nonumber
\\
& & 
\nonumber
\\
& = &
\lim_{N\to\infty}\;
\frac{\lambda^2}{8} \;h^2_\av
\;
\sum_{l=1}^L\,\sum_{k=0}^{L_c}\;
h_k^2 \,\langle\rho_{l,k}\rangle_N
\;+\;
\Oo(\lambda^3)
\nonumber
\end{eqnarray}

\noindent Using (\ref{eq-magicid}), one therefore gets:
 
\begin{theo} 
\label{theo-Lyapsum}
Suppose that the Main hypothesis holds and all channels are elliptic. Then

\begin{equation}
\label{eq-lyapexpsum}
\sum_{l=1}^L\gamma_l
\;= \;
\frac{L\,\lambda^2}{8} \;\left(h^2_\av\right)^2
+\Oo(\lambda^3)
\,.
\end{equation}

\end{theo}

\vspace{.2cm}

Via the Thouless formula \cite{CS}, this theorem also allows to deduce
a $\Oo(\lambda^2)$ correction to the density of states. However, the
term linear in $\lambda$ has to be calculated separately. Theorem 
\ref{theo-Lyapsum} allows to deduce an upper bound on $\gamma_L$,
however, not a very tight one as we shall argue in Section
\ref{sec-speculation}. 

\vspace{.2cm}

\begin{coro} 
Suppose that the Main hypothesis holds and all channels are
elliptic. Then there exists a constant
$c$ such that

$$
\gamma_L
\;\leq\;
c\;\lambda^2 \;\log^2(L)
\;+\;\Oo(\lambda^3)
\,.
$$
\end{coro}

\noindent {\bf Proof}.
Because of the ordering of the Lyapunov exponents, it follows from 
Theorem \ref{theo-Lyapsum} that

$$
\gamma_L\;\leq\;
\frac{\lambda^2}{8} \;\left(h^2_\av\right)^2
\mbox{ . }
$$

\noindent But using $\sin(\eta)\geq \frac{\eta}{\pi}$, one finds
$h^2_\av\leq c_E\,\log(L)$ for some energy dependent constant $c_E>1$.
\hfill $\Box$

\vspace{.2cm}

Finally, let us remark that it is also straight-forward to write out
a perturbative formula for the top Lyapunov exponent:

\begin{equation}
\label{eq-toplyap}
\gamma_1
\;=\;
\lim_{N\to\infty}\;
\frac{\lambda^2}{4}
\left[
\sum_{j=0}^{L_c}
h_\av^2\,h_j^2\;
\langle\rho_{1,j}\rangle_N
-
\frac{1}{2L}
\sum_{j,k=0}^{L_c}h_j^2h_k^2(2-\delta_{j,k})\;
\langle\rho_{1,j}\rho_{1,k}\rangle_N
\right]\;+\;\Oo(\lambda^3)
\,.
\end{equation}

\noindent This will be further analyzed in Section
\ref{sec-speculation}.

\subsection{Weight of elliptic frame vectors in hyperbolic channels}
\label{sec-ellweights}

It follows from Proposition \ref{prop-separ} and the arguments in
its proof that the weight of a elliptic frame vector in the hyperbolic
channels is of order $\Oo(\lambda^2)$, that is for $N$ large enough

$$
\langle \rho_{p,l}\rangle_N\;=\;\Oo(\lambda^2)
\mbox{ , }
\qquad
p=2L_h+2,\ldots,L
\mbox{ , }
\quad
l= 0,\ldots,L_h
\mbox{ . }
$$

\noindent Actually, more detailed information about the leading order
term as well as the redistribution of this weight on the contracting
and expanding basis vectors will be needed below. As it turns out, the
elliptic frame vector is randomly kicked into the hyperbolic channels
and immediately forced back out;
therefore it spends (to leading order) an equal amount of time in the
expanding and contracting hyperbolic directions.

\begin{proposi}
\label{prop-ellweight}
For $p=2L_h+2,\ldots,L$ and $l= 0,\ldots,L_h$, one has

$$
\langle \rho^+_{p,l}\rangle_N
\;=\;
\langle \rho^-_{p,l}\rangle_N
\;+\;
\Oo(N^{-1},\lambda^3)
\mbox{ . }
$$

\end{proposi}

\noindent {\bf Proof.} Let us first calculate $\langle
\rw_l^\sigma|u_p(n+1)\rangle$ in terms of $\langle
\rw_l^\sigma|u_p(n)\rangle$ by using (\ref{eq-action2}). In order to
shorten the appearing expressions, let us drop the argument $n$ in 
$u_p(n)$. The denominator in (\ref{eq-action2})
can be read off

$$
\|\Lambda^p R(1-\lambda P)\,u_1\wedge\ldots\wedge u_p\|^2
\;=\;
\det\left(\langle R(1-\lambda P)u_k| R(1-\lambda P)u_m\rangle_{1\leq
k,m\leq p}\right)
\;=\;
\prod^{L_h}_{k=0}\,e^{2{\eta}_k}+\Oo(\lambda)
\mbox{ . }
$$

\noindent Therefore $\langle
\rw_l^\sigma|u_p(n+1)\rangle$ is equal to

$$
\left(\prod^{L_h}_{k=0}\,e^{-2{\eta}_k}\right)\;
\langle
\Lambda^{p-1} R(1-\lambda P)\,u_1\wedge\ldots\wedge u_{p-1}\wedge \rw_l^\sigma|
\Lambda^p R(1-\lambda P)\,u_1\wedge\ldots\wedge u_p
\rangle
\,(1+\Oo(\lambda))
.
$$

\noindent The appearing scalar product in $\Lambda^p\RR^{2L}$ is given by

$$
\det
\left(
\begin{array}{cc}
\langle R(1-\lambda P)u_k| R(1-\lambda P)u_m\rangle_{1\leq
k,m\leq p-1} & \langle R(1-\lambda P)u_k| R(1-\lambda P)u_p\rangle_{1\leq
k\leq p-1}
\\
&
\\
\langle \rw^\sigma_l| R(1-\lambda P)u_m\rangle_{1\leq
m\leq p-1}
&
\langle \rw^\sigma_l| R(1-\lambda P)u_p\rangle
\end{array}
\right)
\;.
$$

\noindent All the off-diagonal matrix elements
are $\Oo(\lambda)$ due to the results of Section \ref{sec-separ}, except
when $\sigma=+$. 
In the latter case, the entries of the lower left corner
are $\Oo(1)$ for $m=2l,2l+1$ (again by Proposition
\ref{prop-separ}). Hence the contributions to the determinant up to
order $\Oo(\lambda)$ are given by the product of the diagonal elements
and (in the case $\sigma=+$) by two transpositions. The diagonal
elements of the upper left part are treated as above and cancel with
the denominator. Therefore,

\begin{eqnarray}
& & \!\!\!\!\!\!\!\!\!\langle
\rw_l^\sigma|u_p(n+1)\rangle
\;=\;
e^{\sigma\eta_l}\;
\langle
\rw_l^\sigma|(1-\lambda P)u_p(n)\rangle
\nonumber
\\
& &
\nonumber
\\
& &
\;\;\;\;\;\;\;\;\;\;\;-\;
\delta_{\sigma,+}
e^{-2\eta_l}
\sum_{j=0,1}
\;\langle
\rw_l^+|R(1-\lambda P)u_{2l+j}\rangle
\langle
R(1-\lambda P)u_{2l+j}|R(1-\lambda P)u_p\rangle
\;+\;\Oo(\lambda^2)
\mbox{ . }
\nonumber
\end{eqnarray}

\noindent For the case
$\rw^-_l$, one reads off 

\begin{equation}
\langle
\rw_l^-|u_p(n+1)\rangle
\;=\;e^{-\eta_l}\;\langle
\rw_l^-|(1-\lambda P)u_p(n)\rangle
\;+\;\Oo(\lambda^2)
\mbox{ , }
\label{eq-hist1}
\end{equation}

\noindent while a bit of algebra invoking again Proposition
\ref{prop-separ} shows

\begin{equation}
\langle
\rw_l^+|u_p(n+1)\rangle
\;=\;e^{-\eta_l}\;\langle
\rw_l^+|(1+\lambda P^*)u_p(n)\rangle
\;+\;\Oo(\lambda^2)
\mbox{ , }
\label{eq-hist2}
\end{equation}

Now set $J^\sigma(N)=\EE\,\frac{1}{N}\sum_{n=0}^{N-1}
\,|\langle \rw_l^\sigma|u_p(n)\rangle|^2$. Because $\EE\, P=0$, one gets by
going back in history once

\begin{eqnarray}
J^-(N) & = &
e^{-2\eta_l}\,J^-(N) +\lambda^2\,e^{-2\eta_l}\,
\EE\,\frac{1}{N}\sum_{n=0}^{N-1}
\,|\langle \rw_q^-|P|u_p(n)\rangle|^2
\;+\;\Oo(N^{-1},\lambda^3)
\nonumber
\\
& = &
\lambda^2\,\frac{1}{e^{2\eta_l}-1}\,
\EE\,\frac{1}{N}\sum_{n=0}^{N-1}
\,\langle u_p(n)|P^*|\rw_l^-\rangle\langle \rw_l^-|P|u_p(n)\rangle
\;+\;\Oo(N^{-1},\lambda^3)
\mbox{ . }
\nonumber
\end{eqnarray}

\noindent The appearing oscillatory sum can be treated by the same
argument as in the proof of Lemma \ref{lem-osci}(i). This gives

$$
J^-(N)
\;=\;
\lambda^2\,\frac{1}{4L}\,\frac{1}{e^{2\eta_l}-1}\;
h_l^2\,\sum_{k=0}^{L_c}h_k^2\;\langle\rho_{p,k}\rangle_N
\;+\;\Oo(N^{-1},\lambda^3)
\mbox{ . }
$$

\noindent This
argument can be repeated for $J^+(N)$ using
(\ref{eq-hist2}) instead of (\ref{eq-hist1}). One finds
$J^+(N)=J^-(N)+\Oo(N^{-1},\lambda^3)$. Now the whole argument can be
repeated for $\rw_{L-l}^\sigma$. Finally summing the
contributions of $\rw_l^\sigma$ and  $\rw_{L-l}^\sigma$ allows to 
conclude the
proof. 
\hfill $\Box$

\subsection{Perturbative formula for bottom Lyapunov exponent}
\label{sec-perturbative}

The aim is now to generalize the perturbative calculation of the
Lyapunov exponents given in Section \ref{sec-lyapsymp} to the case
where there are both elliptic and hyperbolic channels. It is
convenient to introduced the scaled frame vectors:

$$
\hat{u}_l(n)
\;=\;
e^{-\hat{\eta}_l}\;{u}_l(n)
\mbox{ , }
\qquad
\hat{\eta}_l
\;=\;
\frac{1}{2}(1-g_{[\frac{l}{2}]}^2)\eta_{[\frac{l}{2}]}
\mbox{ . }
$$

\noindent Note that, while the index on $\eta_l$ matches the channel
index, the one on  $\hat{\eta}_l$ matches the frame vector: for a
hyperbolic frame vector, $\hat{\eta}_l$ is the expansion exponent in
the direction into which $u_l$ is alined by Proposition
\ref{prop-separ}, but for an elliptic frame vector $u_l$, one has 
$\hat{\eta}_l=0$. 
Using the multilinearity of the determinant and then 
$\log\det_p=\mbox{Tr}_p\log$, one finds

$$
\sum_{l=1}^p\,\gamma_l\,-\,\hat{\eta}_l
\;=\;
\lim_{N\to\infty}\,\frac{1}{2N}\;
\sum_{n=0}^{N-1}\;
\EE\;
\mbox{Tr}_p\left(\,\log
\left(\langle \hat{u}_l(n)|\;
|R(1-\lambda P)|^2\,|
\hat{u}_k(n)\rangle_{1\leq l,k\leq p}\right)
\,\right)
\mbox{ . }
$$

\noindent The matrix elements of the 
leading order $|R(1-\lambda P)|^2=R^*R+\Oo(\lambda)$ 
now give a unit matrix ${\bf 1}_p$. 
In fact, using the orthonormality property of the frame
vectors and  inserting
(\ref{eq-decomp}), one finds

\begin{equation}
\langle\hat{u}_l(n)|R^*R|\hat{u}_k(n)\rangle-\delta_{l,k}
\;=\;
\sum_{m=0}^{L_c}\sum_{\sigma=\pm}
\left(
e^{-\hat{\eta}_l-\hat{\eta}_k}\,e^{\sigma\eta_m(1-g_m^2)}
-1\right)\,
\langle u_l(n)|\pi_m^\sigma|u_k(n)\rangle
\;=\;
\Oo(\lambda)
\mbox{ . }
\nonumber
\end{equation}

\noindent Moreover, if $l=k$ this expression is $\Oo(\lambda^2)$.
Indeed, for a hyperbolic frame vector $u_l$, the summand $m=l$ has a
vanishing prefactor and all the others are $\Oo(\lambda^2)$
by Proposition \ref{prop-separ}, while for elliptic $u_l$, all $m$
corresponding to elliptic channels have vanishing prefactors and all
the remaining hyperbolic $m$ are $\Oo(\lambda^2)$
by Proposition \ref{prop-separ} (actually, they were even calculated
in Proposition \ref{prop-ellweight}).
Now around  ${\bf 1}_p$ 
the logarithm can be expanded. In the expansion, the terms
linear in $P$ can be discarded because $\EE\,P=0$. 
Hence $\sum_{l=1}^p\gamma_l-\hat{\eta}_l$ is up to
$\Oo(\lambda^3)$ equal to

\begin{eqnarray}
\lim_{N\to\infty}\frac{1}{2N}
\sum_{n=0}^{N-1}
\!\!\!\! & \EE & \!\!\! \!\left[ 
\sum_{l=1}^p
(\langle\hat{u}_l(n)|R^*R|\hat{u}_l(n)\rangle-1)
\;-\;\frac{1}{2}
\sum_{l,k=1,\,l\neq k}^p
\,\langle\hat{u}_l(n)|R^*R|\hat{u}_k(n)\rangle^2
\right.
\nonumber
\\
& &
\nonumber
\\
& &
+\;\lambda^2
\sum_{l=1}^p
\langle\hat{u}_l(n)|\,|RP|^2\,|\hat{u}_l(n)\rangle
\left.
\;-\;\frac{\lambda^2}{2}
\sum_{l,k=1}^p
\langle\hat{u}_l(n)|(R^*RP+P^*R^*R)|\hat{u}_k(n)\rangle^2
\right]
\,.
\nonumber
\end{eqnarray}

\noindent When there are no hyperbolic channels, $R$ is orthogonal and
the formula
reduces to (\ref{eq-lyapexpan}). The bottom exponent can now be
obtained by substraction. Let us suppose that $\hat{\eta}_L=0$ which
means that $E$ is in the spectrum of $H_L(0)$:

\begin{eqnarray}
\!\!\!\! \!\!\!\! \!\!\!\! \!\!
\gamma_L
\;=\;
\left.
\lim_{N\to\infty}\,\frac{1}{N}\;
\sum_{n=0}^{N-1}\;
\EE \right[ 
& & 
\!\!\!\! \!\!\!\! \!\!\frac{1}{2}\;
\left(\langle {u}_L(n)|R^*R|{u}_L(n)\rangle-1\right)
\label{eq-expanbot1}
\\
& & \!\!\!\! \!\!
-\;\frac{1}{2}\;
\sum_{l=1}^{L-1}\;
\langle{u}_L(n)|R^*R|\hat{u}_l(n)\rangle^2
\label{eq-expanbot2}
\\
& & \!\!\!\! \!\!
+\;\frac{\lambda^2}{2}\;
\langle{u}_L(n)|\,|RP|^2\,|{u}_L(n)\rangle
\label{eq-expanbot3}
\\
& & \!\!\!\! \!\!
+\;\frac{\lambda^2}{4}
\;
\langle{u}_L(n)|(R^*RP+P^*R^*R)|{u}_L(n)\rangle^2
\label{eq-expanbot4}
\\
& & \!\!\!\! \!\!
\left.
-\;\frac{\lambda^2}{2}\;
\sum_{l=1}^L\;
\langle{u}_L(n)|(R^*RP+P^*R^*R)|\hat{u}_l(n)\rangle^2
\;\right]
\;+\;\Oo(\lambda^3)
\,.
\label{eq-expanbot5}
\end{eqnarray}

Now the terms (\ref{eq-expanbot1}) to (\ref{eq-expanbot5}), each by
definition 
containing the average 
$\frac{1}{N}
\sum_{n=0}^{N-1}\,
\EE$ but not the limit $N\to\infty$, will be
treated separately.
Inserting (\ref{eq-decomp}) and using the normalization property
(\ref{eq-magicid}),

$$
(\ref{eq-expanbot1})
\;=\;
\frac{1}{2}\;
\sum_{l=0}^{L_c}
\;
\left(e^{(1-g_l^2)\eta_l}-1\right)\;
\langle\rho_{L,l}^+\rangle_N
+
\left(e^{-(1-g_l^2)\eta_l}-1\right)\;
\langle\rho_{L,l}^-\rangle_N
\;=\;
\sum_{l=0}^{L_h}
\;
(\cosh(2\eta_l)-1)\;
\langle\rho_{L,l}^+\rangle_N
\mbox{ , }
$$

\noindent the second step because
the appearing averaged weights are equal by Proposition
\ref{prop-ellweight}. Next, using orthogonality of $u_l$ and $u_L$,

$$
(\ref{eq-expanbot2})
\;=\;
-
\frac{1}{2N}\;
\sum_{n=0}^{N-1}\;
\sum_{l=1}^{L-1}\;
\EE
\left(
\sum_{k=0}^{L_c}\sum_{\sigma=\pm}
\;
\left(e^{\sigma(1-g_k^2)\eta_k}-1\right)\;
\langle u_L(n)|\pi^\sigma_k|\hat{u}_l(n)\rangle
\right)^2
\mbox{ . }
$$

\noindent Again, the sum over $k$ is actually restricted to the hyperbolic
channels. But for hyperbolic a channel $k$, one has 
$\langle u_L(n)|\pi^\sigma_k|\hat{u}_l(n)\rangle
=\Oo(\lambda^2)$ unless $\sigma=+$ and $l=2k,2k+1$ by Proposition
\ref{prop-separ}. Hence 

$$
(\ref{eq-expanbot2})
\;=\;
-
\frac{1}{2}\;
\sum_{k=0}^{L_h}\;\left(e^{2\eta_k}-1\right)^2e^{-2\eta_k}
\langle\rho_{L,k}^+\rangle_N
\;+\;
\Oo(\lambda^4)
\mbox{ , }
$$

\noindent which shows that to leading order
(\ref{eq-expanbot1}) and (\ref{eq-expanbot2}) compensate.
The contribution (\ref{eq-expanbot3}) was already calculated in Lemma
\ref{lem-osci}(i). It will be compensated by (\ref{eq-expanbot5})
which is a bit more
cumbersome to treat. Hence let us formulate it as a separate lemma.

\begin{lemma} 
\label{lem-osci2} Let $p>2L_h+1$ and suppose that the Main hypothesis holds.

$$
\frac{1}{N}\sum_{n=0}^{N-1}\;
\EE
\,\sum_{l=1}^L
\langle u_p(n)|(R^*RP+P^*R^*R)|\hat{u}_l(n)\rangle^2
\;=\;
\frac{1}{2}\,
h_\av^2
\;
\sum_{l=0}^{L_c}
h_l^2\;
\langle
\rho_{p,l}
\rangle_N
\;+\;\Oo(N^{-1},\lambda)
\;.
$$

\end{lemma}

\noindent {\bf Proof.} Let us call the l.h.s. $J(N)$. 
Because
$R^*R|u_k(n)\rangle=e^{2\hat{\eta}_k}|u_k(n)\rangle+\Oo(\lambda)$,  

$$
J(N)
\;=\;
\frac{1}{N}\sum_{n=0}^{N-1}\;
\EE
\,
\sum_{l=1}^L
\langle u_p(n)|
(e^{-\hat{\eta}_l}P+e^{\hat{\eta}_l}P^*)
|u_l(n)\rangle
\;
\langle u_l(n)|
(e^{\hat{\eta}_l}P+e^{-\hat{\eta}_l}P^*)
|u_p(n)\rangle
\;+\;\Oo(\lambda)
\mbox{ . }
$$

\noindent Let us split $J(N)=J_h(N)+J_e(N)$ where $J_h(N)$ contains
the sum over indices $l=1,\ldots,2L_h+1$ corresponding to hyperbolic
frame vectors and $J_e(N)$ the remainder corresponding to elliptic frame
vectors. In $J_h(N)$, one can replace up to $\Oo(\lambda)$

$$
|u_1(n)\rangle\langle u_1(n)|\;=\;
\pi_0^+
\mbox{ , }
\qquad
\sum_{j=0,1}|u_{2k+j}(n)\rangle\langle u_{2k+j}(n)|\;=\;
\pi_j^+
\mbox{ , }
$$

\noindent by Proposition \ref{prop-separ}. Hence

$$
J_h(N)
\;=\;
\frac{1}{N}\sum_{n=0}^{N-1}\;
\EE
\,
\sum_{k=0}^{L_h}
\langle u_p(n)|
(e^{-{\eta}_k}P+e^{{\eta}_k}P^*)
\,\pi_k^+\,
(e^{{\eta}_k}P+e^{-{\eta}_k}P^*)
|u_p(n)\rangle
\;+\;\Oo(\lambda)
\mbox{ , }
$$

\noindent and an oscillatory sum argument implies that $J_h(N)$ is
equal to

$$
\sum_{m=0}^{L_c}\sum_{\sigma=\pm}\;
\langle \rho_{p,m}^\sigma\rangle_N
\sum_{k=0}^{L_h}
\;\EE
\,
\langle w_m^\sigma|
(e^{-{\eta}_k}P+e^{{\eta}_k}P^*)
\,\pi_k^+\,
(e^{{\eta}_k}P+e^{-{\eta}_k}P^*)
| w_m^\sigma\rangle
\;+\;\Oo(N^{-1},\lambda)
\mbox{ , }
$$

\noindent where $w_m^\sigma$ is some unit vector satisfying 
$\pi_m^\sigma w_m^\sigma=w_m^\sigma$.
The expectation value of the
last factor is by Lemma
\ref{lem-matrixel} independent of $w_m^\sigma$. 
Now $\langle \rho_{p,m}^+\rangle_N
=\langle \rho_{p,m}^-\rangle_N+\Oo(\lambda)$
by Proposition \ref{prop-ellweight} and (\ref{eq-ellipweight}).
Of the four terms, the
$(P,P)$ and $(P^*,P^*)$ pairs vanish after summing over $\sigma$ because of
the sign in Lemma \ref{lem-matrixel}(ii). The remaining terms 
$(P,P^*)$ and $(P^*,P)$ are given by Lemma \ref{lem-matrixel}(iii) so that

$$
J_h(N)
\;=\;
\frac{1}{2}\;
\sum_{m=0}^{L_c}\,h_m^2
\,\langle \rho_{p,m}^\sigma\rangle_N
\;\frac{1}{L}\,
\sum_{k=0}^{L_h}
\;
h_k^2\,\nu_k\,\cosh(2\eta_k)
\;+\;\Oo(N^{-1},\lambda)
\mbox{ . }
$$

Finally for the $l$ in the sum of $J_e(h)$, 
one has $e^{\hat{\eta}_l}P+e^{-\hat{\eta}_l}P^*=\tilde{P}$. Therefore,
this is actually the term treated in Lemma
\ref{lem-osci}(iii). Combining proves the lemma.
\hfill $\Box$

\vspace{.2cm}

Therefore, (\ref{eq-expanbot3}) and (\ref{eq-expanbot5}) 
compensate to leading order just as do
(\ref{eq-expanbot1}) to (\ref{eq-expanbot2}). The
leading order contribution to 
$\gamma_L$ is thus solely given by (\ref{eq-expanbot4}).
Noting that
$R^*R|u_L(n)\rangle=|u_L(n)\rangle+\Oo(\lambda)$, 
the contribution (\ref{eq-expanbot4}) was already dealt with, in
Lemma \ref{lem-osci}(ii) and we have proven:

\vspace{.2cm}

\begin{theo} 
\label{theo-Lyapasymp}
Suppose that $E\in\RR$ 
is in the spectrum of $H_L(0)$ and satisfies the Main 
hypothesis. Then

\begin{equation}
\label{eq-lyapexpansympgeneral}
\gamma_L
\;= \;
\lim_{N\to\infty}\;
\frac{\lambda^2}{8L}
\sum_{j,k=0}^{L_c}h_j^2h_k^2
(2-\delta_{j,k})\;
\langle\rho_{L,j}\rho_{L,k}\rangle_N
\;+\;\Oo(\lambda^3)
\,.
\end{equation}

\end{theo}

\vspace{.2cm}

The presented techniques also allow to write out formulas
for the top Lyapunov exponent and the sum of the positive Lyapunov
exponents.

\vspace{.2cm}

\noindent {\bf Proof} of Theorem \ref{theo-main}. (i) follows
immediately from
$h_k^2\geq 1$ and the fact that $\langle\rho_{L,j}\rho_{L,k}\rangle_N$
is a probability distribution.
(ii) For hyperbolic channels $j,k=0,\ldots L_h$, the weights in 
(\ref{eq-lyapexpansympgeneral}) are $\Oo(\lambda)$ by Proposition
\ref{prop-separ} so that they can be
neglected. For the remaining elliptic channels $j$, one has
$h_j^2\geq h_{L_c}^2$. But
$h_{L_c}^2=1/\sin(\eta_{L_c})=
(1-\frac{\mu_{L_c}^2}{4})^{-1/2}=
(\epsilon-\frac{\epsilon^2}{4})^{-1/2}\geq \epsilon^{-1/2}$.
Replacing this concludes the proof.
\hfill $\Box$

\section{More insights on the channel weights}
\label{sec-speculation}

This short section does not contain rigorous results. The aim is to
get a better understanding of
the averaged channel weights entering in the perturbative
formulas above. Again, for simplicity, let us restrict ourselves to
the situation where there are only elliptic channels. We first focus
on the weights of the first frame vector $u_1$. It follows from 
(\ref{eq-action}) that, up to $\Oo(\lambda^3)$,

\begin{eqnarray}
\EE\,\rho_{1,k}(n+1)- \rho_{1,k}(n)
& = & \lambda^2 \;\EE
\Big[
\langle u_1|P^*\pi_k P|u_1\rangle
-
\langle u_1|\pi_k|u_1\rangle
\langle u_1|\,|P|^2\,|u_1\rangle
\nonumber
\\
& & \;\;\;\;\;\;\;\; +
\left.
\langle u_1|\pi_k|u_1\rangle
\langle u_1|\tilde{P}|u_1\rangle^2
-
\langle u_1|(P^*\pi_k+\pi_kP)|u_1\rangle
\langle u_1|\tilde{P}|u_1\rangle
\right]
\;,
\nonumber
\end{eqnarray}

\noindent where the index $n$ is left out on the r.h.s. and the
expectation values is over $P(n+1)$ only.
Now let us average $\EE\,\frac{1}{N}\sum_{n=0}^{N-1}$ and suppose that
the limit exists. Then the l.h.s. vanishes. Hence the coefficient of
$\lambda^2$ on the r.h.s. has to vanish as well, up to $\Oo(\lambda)$.
Calculating it with
an oscillatory sum argument as in Section \ref{sec-osci}
shows that for all $k=0,\ldots, L_c$:

\begin{eqnarray}
0 & = &
\frac{1}{2L}\;
\sum_{l=0}^{L_c}\nu_k h_k^2h_l^2\;\langle \rho_{1,l}\rangle
-
\frac{1}{2L}\;
\sum_{l,m=0}^{L_c}\nu_m h_m^2h_l^2
\langle \rho_{1,l}\rho_{1,k}\rangle
\nonumber
\\
& & +\;
\frac{1}{2L}\sum_{l,m=0}^{L_c}
h_l^2h_m^2(2-\delta_{l,m})\;
\langle\rho_{1,l}\rho_{1,m}\rho_{1,k}\rangle
\;-\;
\frac{1}{2L}\sum_{l=0}^{L_c}
h_l^2h_k^2(2-\delta_{k,l})\;\langle\rho_{1,l}\rho_{1,k}\rangle
\,.
\nonumber
\end{eqnarray}

\noindent These equations give relations between the averaged first,
second and third moments of the weights
$\rho_{1,l}$. Analogously, one can write out
equations for $\langle
\rho_{1,l}\rho_{1,k}\rangle$ which then invoke up to the averaged 
sixth moments of the channel weights, and so on. This gives a
hierarchy of equations for the channel weights. 
It results that the weights are independent
of $\lambda$ and only depend on energy $E$ (through the 
the frequencies $\eta$).

\vspace{.2cm}

It order to analyse these equations, let us close them
already at first order by assuming factorization
$\langle\rho_{1,l}\rho_{1,k}\rangle=
\langle\rho_{1,l}\rangle\,\langle\rho_{1,k}\rangle$ and 
$\langle\rho_{1,l}\rho_{1,m}\rho_{1,k}\rangle=
\langle\rho_{1,l}\rangle\,
\langle\rho_{1,m}\rangle\,\langle\rho_{1,k}\rangle$. 
Furthermore we
neglect the $\delta_{m,l}$ and suppose $\nu_k=2$,
both approximations
which are $\Oo(1/L)$ w.r.t. the other
terms.
Now the sum over $l$ factors and one
obtains:

$$
0 \; = \;
h_k^2\;-\;\langle\rho_{1,k}\rangle\,\sum_{m=0}^{L_c}
h_m^2
\;+\;
\langle\rho_{1,k}\rangle
\sum_{m=0}^{L_c}
h_m^2\;\langle\rho_{1,m}\rangle
\;-\;
h_k^2\;
\langle\rho_{1,k}\rangle
\;.
$$

\noindent These equations have the unique solution (recall
$h_k^2=1/\sin(\eta_k)$) 

$$
\langle\rho_{1,k}\rangle
\;=\;
\frac{1}{1+Z\,\sin(\eta_k)}
\mbox{ , }
$$

\noindent where $Z\geq 0$ is such that normalization
$\sum_{k=0}^{L_c}\langle\rho_{1,k}\rangle=1$ is assured. One easily
verifies that $Z\sim L$. For small $k$ (and large ones $L-k$ as well) 
one has $\sin(\eta_k)\sim k/L$. Hence the weight on these channels is
of order of unity, while it is of order $1/L$ on the others. But the
channels with small and large $k$ are precisely those near the band
edges in Fig.~1(ii) where the rotation frequency is small. Hence the
weight of $u_1$ is concentrated on the slowly rotating channels for
which $h_k^2=\Oo(L)$ (they
can be considered to be most similar to hyperbolic channels). 
Hence it is expected from (\ref{eq-toplyap}) that
$\gamma_1=c\,\lambda^2+\Oo(\lambda^2)$ where $c=\Oo(1)$ as
$L\to\infty$. Presumably, only the first few exponents are
considerably larger than $\gamma_L$.

\vspace{.2cm}

Due to symplectic blocking, the weight of $u_2$ has to be centered on
slightly faster rotating channels, ecc\`etera. In conclusion, the weight of
last frame vector $u_L$ is expected to be concentrated on the channels
which rotate the fastest and hence correspond to the band center. In these
channels, $h_k^2=\Oo(1)$ unless $E$ is an internal band edge in which case
$h_h^2=\infty$. Therefore, away from these points the lower bound $h_k^2\geq 1$
which allowed to deduce Theorem \ref{theo-main} from Theorem
\ref{theo-Lyapasymp} is presumably not so bad because the weights of
$u_L$ enter into formula (\ref{eq-lyapexpansympgeneral}). In the case of
mixed elliptic and hyperbolic channels, we expect the above argument to
hold within the elliptic part of $\CC^{2L}$, namely for the weight vectors
not alined to the hyperbolic channels by Proposition \ref{prop-separ}.

\vspace{.4cm}

\noindent {\bf Acknowledgments:} This work profited from financial
support of the SFB 288. While this paper was with the referees, we have done
in collaboration with R. R\"omer extensive numerical studies of the
perturbative formula (\ref{eq-lyapexpansympgeneral}). 
It very well reproduces the energy dependence of
the Lyapunov exponent (as calculated with the standard transfer matrix
method), and this even for surprisingly large disorder strengths.


\end{document}